\documentclass[11pt]{article}
\usepackage{a4,bm,epsfig,booktabs}
\usepackage{setspace}
\newcommand{\laufj}{\kappa}
\newcommand{\laufm}{m}
\newcommand{\restkonst}{T}
\newcommand{\rest}{c}
\newcommand{\Rest}{F}
\newcommand{\Restkonst}{C}

\makeatletter
\newcounter{tab}
\newcommand{\fusszeile}{}
                        {\par\noindent\fusszeile
                         \end{center}}
\makeatother
\usepackage{ifthen,array}
\newcommand{\ams}{\usepackage{amsfonts,amssymb,amsmath}}

\ams
\allowdisplaybreaks[3]
\newlength{\textwidthorig}
\newlength{\oddsidemarginorig}
\newlength{\textheightorig}
\newlength{\topmarginorig}
\def\seitenlaengenabsolut#1 #2 #3 #4 {\setlength{\textwidth}{#1}
                                      \setlength{\oddsidemargin}{#2}
                                      \setlength{\textheight}{#3}
                                      \setlength{\topmargin}{#4}}
\def\seitenlaengenrelzustandard#1 #2 #3 #4 {\setlength{\textwidth}{\textwidthorig+#1}
                                            \setlength{\oddsidemargin}{\oddsidemarginorig+#2}
                                            \setlength{\textheight}{\textheightorig+#3}
                                            \setlength{\topmargin}{\topmarginorig+#4}}
\def\seitenlaengenrelzuvorher#1 #2 #3 #4 {\addtolength{\textwidth}{#1}
                                          \addtolength{\oddsidemargin}{#2}
                                          \addtolength{\textheight}{#3}
                                          \addtolength{\topmargin}{#4}}
\newcommand{\standardseite}{\seitenlaengenrelzuvorher2.2cm -0.8cm 1.8cm -1.5cm }   %
\standardseite
\newcommand{\leerezeile}{\vspace{2ex}}
\newlength{\laengespatium}

\newcommand{\nach}{\longrightarrow}      

\newcommand{\auf}{\longmapsto}           
\newcommand{\txtauf}[1]{\auf}            
\newcommand{\impliz}{\Longrightarrow}    
\newcommand{\aequ}{\Longleftrightarrow}  
 
\newcommand{\invimpliz}{\Longleftarrow}  
\newcommand{\gegen}{\rightarrow}         
\newcommand{\iso}{\cong}                 
\newcommand{\ident}{\equiv}              
\newcommand{\teilmenge}{\subseteq}       
\newcommand{\obermenge}{\supseteq}       
\newcommand{\aeqrel}{\sim}               

\newcommand{\senk}{\perp}                

\newcommand{\kreuz}{\times}              

\newcommand{\einschr}[1]{{}\arrowvert_{#1}}      
\newcommand{\dirsum}{\oplus}           

\newcommand{\betraganpass}[1]%
           {\left| #1 \right|}           
\newcommand{\bigbetrag}[1]%
           {\bigl|{#1}\bigr|}            
\newcommand{\betrag}[1]%
           {|{#1}|}                      
\newcommand{\betragnichtanpass}[1]%
           {\mid #1 \mid}                
\newcommand{\norm}[1]%
           {{}{\parallel}#1{\parallel}{}}      
\newcommand{\erww}[1]%
           {\langle #1 \rangle}          
\newcommand{\skalprod}[2]%
           {\langle #1,#2 \rangle}       
\newcommand{\quer}{\overline}            
\newcommand{\inv}[1]{\frac{1}{#1}}       
\newcommand{\re}{\text{Re }}                           
\newcommand{\im}{\text{im\;}}                          
\newcommand{\tr}{\text{tr}}                           
\newcommand{\diag}{\text{diag }}                       
\newcommand{\dd}{\text{d}}                             
\newcommand{\e}{\text{e}}                              
\newcommand{\NULL}{\mathbf{0}}                         
\newcommand{\EINS}{{\boldsymbol{1}}}                   
\newcommand{\field}[1]{\mathbb{#1}}                    
\newcommand{\C}{{\field{C}}}                           
\newcommand{\N}{{\field{N}}}                           
\newcommand{\R}{{\field{R}}}                           
\newcommand{\gl}{\mathfrak{gl}}                        
\newcommand{\rnkl}[2]{\raisebox{-0.4ex}{$#1$}%
\raisebox{-0.12ex}{{\large$\setminus$}}\,#2}   
\newcommand{\agb}{{\overline{{\cal A}/{\cal G}}}}      
\newcommand{\agbfact}[1][]{\text{$\agb/\!\aeqrel$}}    
\newcommand{\Gb}{{\overline{{\cal G}}}}                

\newcommand{\qa}{{\quer{A}}}                           

\newcommand{\holgr}{{\mathbf H}}                       
\newcommand{\bz}{{\mathbf B}}                          

\newcommand{\LG}{{\mathbf{G}}}                         
\newcommand{\aeqrelzush}[1][]{\sim}                    

\newcommand{\nklza}[1][]{\ifthenelse{\equal{#1}{}}     
                                    {\rnkl{Z(\holgr_\qa)}{\LG}}        
                                   {\rnkl{Z(\holgr_{#1})}{\LG}}}       
\newcommand{\nkla}[1][]{\ifthenelse{\equal{#1}{}}      
                                    {\rnkl{\bz(\qa)}{\Gb}}        
                                    {\rnkl{\bz(#1)}{\Gb}}}       

\newcommand{\YM}{{\text{YM}}}                          

\newcommand{\ymwirk}[1][]{\ifthenelse{\equal{#1}{}}{S_{\YM}}{S_{\YM,#1}}}

\newcommand{\bmat}{\begin{pmatrix}}
\newcommand{\emat}{\end{pmatrix}}

\newcommand{\ListNullAbstaende}{\setlength{\topsep}{0pt}%
                                \setlength{\parskip}{0pt}%
                                \setlength{\partopsep}{0pt}%
                                \setlength{\itemsep}{0pt}%
                                \setlength{\parsep}{0pt}}
\newcommand{\ListNurAnstrichAbstand}{\setlength{\topsep}{0pt}%
                                     \setlength{\parskip}{0pt}%
                                     \setlength{\partopsep}{0pt}%
                                     \setlength{\parsep}{0pt}}
\newenvironment{StandardListe}[2]%
               {\begin{list}%
                      {#1}%
                      {\settowidth{\leftmargin}{M#1}%
                       \settowidth{\labelwidth}{#1}%
                       \settowidth{\labelsep}{M}%
                       #2%
                      }%
                }%
               {\end{list}}%
\newenvironment{EinfachListe}[1]%
               {\begin{StandardListe}{#1}{\ListNullAbstaende}}%
               {\end{StandardListe}}%
               {\begin{StandardListe}{#1}{\ListNurAnstrichAbstand}}%
               {\end{StandardListe}}%
\newcommand{\labelsatz}[1]{#1}
\newcounter{listennr}                      %
\newlength{\hilfslaenge}
\newlength{\stdlabellaenge}
\newlength{\maximum}
\newcommand{\stdlabel}{}
\newcommand{\Maximum}{}
\newcommand{\iitem}[1][]{\ifthenelse{\equal{#1}{}}%
                           {\item \setlength{\hilfslaenge}{\stdlabellaenge}}%
                           {\item[\labelsatz{#1}\hfill]%
                            \settowidth{\hilfslaenge}{\labelsatz{#1}}}%
                         \ifthenelse{\lengthtest{\maximum < \hilfslaenge}}%
                           {\setlength{\maximum}{\hilfslaenge}%
                            \ifthenelse{\equal{#1}{}}%
                               {\renewcommand{\Maximum}{\stdlabel}}%
                               {\renewcommand{\Maximum}{#1}}}%
                           {}%
                      }      
\makeatletter
\newenvironment{AutoLabelLaengenListe}[2][]%
               {\begin{list}%
                      {\labelsatz{#1}\hfill}%
                      {\stepcounter{listennr}%
                       \settowidth{\leftmargin}{M\labelsatz{\ref{listnr\arabic{listennr}}}}%
                       \settowidth{\labelwidth}{\labelsatz{\ref{listnr\arabic{listennr}}}}%
                       \settowidth{\labelsep}{M}%
                       \settowidth{\stdlabellaenge}{\labelsatz{#1}}%
                       \renewcommand{\stdlabel}{#1}%
                       #2%
                       \renewcommand{\Maximum}{}%
                      }%
                }%
               {\renewcommand{\@currentlabel}{\Maximum}%
                \label{listnr\arabic{listennr}}%
                \end{list}%
                }%
\makeatother
\newenvironment{StandardEinrueckung}[2]%
               {\begin{list}%
                      {#1}%
                      {\settowidth{\leftmargin}{M#1}%
                       \settowidth{\labelwidth}{#1}%
                       \settowidth{\labelsep}{M}%
                       #2%
                      }%
                \item}%
               {\end{list}}%
\newenvironment{Einrueckungpur}[1]%
               {\begin{StandardEinrueckung}{#1}{\ListNullAbstaende}}%
               {\end{StandardEinrueckung}}%
\newenvironment{Einrueckung}[1]%
               {\begin{StandardEinrueckung}{#1}{\setlength{\parsep}{0pt}}}%
               {\end{StandardEinrueckung}}%

\makeatletter
\newcommand{\EineNumZeileGleichung}[2][0.5ex]
           {
            
            \vspace{#1} 
            \noindent
            \stepcounter{equation}
            \renewcommand{\@currentlabel}{\arabic{equation}}%
            \phantom{(\arabic{equation})}\hspace*{\fill}
            $\displaystyle{#2}$
            \hspace*{\fill}
            (\arabic{equation})

            \vspace{#1} 
            
           }
\makeatother
\makeatletter
\newcommand{\EineErwNumZeileGleichung}[2][0.5ex]
           {
            
            \vspace{#1} 
            \noindent
            \stepcounter{equation}
            \renewcommand{\@currentlabel}{\arabic{equation}}%
            \phantom{(\arabic{equation})}\hspace*{\fill}
            #2 %
            \hspace*{\fill}
            (\arabic{equation})

            \vspace{#1} 
            
           }
\makeatother
\newcommand{\breitrel}[1]{\hspace*{\tabcolsep} #1 \hspace*{\tabcolsep}}
\newlength{\abstaug}              %
\newenvironment{AllgUnnumGleichung}[2][1.0ex]
               {
  
                \setlength{\abstaug}{#1}
                \vspace{\abstaug}
                \hspace*{\fill}
                $\begin{array}[t]{#2}
                }%
               {\end{array}$
                \hspace*{\fill}
  
                \vspace{\abstaug}

                }%
\newenvironment{AllgNumGleichung}[2][0.0ex]
               {
  
                \setlength{\abstaug}{#1}
                \vspace{\abstaug}
                $\begin{tabular*}{\textwidth}[t]{#2}
                }%
               {\end{tabular*}$

                \vspace{\abstaug}

               }%
\newenvironment{StandardUnnumGleichungKlein}[1][0ex]
               {\renewcommand{\s}{\\[#1] }%
                \begin{AllgUnnumGleichung}{rcl}}%
               {\end{AllgUnnumGleichung}}%
\newcommand{\s}{\\[0ex] }             %
\newenvironment{StandardUnnumGleichung}[1][0ex]%
               {\renewcommand{\s}{\\[#1] }%
                \begin{AllgUnnumGleichung}{>{\displaystyle}rc>{\displaystyle}l}}%
               {\end{AllgUnnumGleichung}}%
\newenvironment{XrelYZNumGleichung}[1][0ex]
               {\renewcommand{\s}{\\[#1] }%
                \begin{AllgNumGleichung}{rcll}}%
               {\end{AllgNumGleichung}}%
\newcommand{\erl}[1]{\hfill\mbox{\hspace*{1.5em}\small (#1)}}

\newcommand{\erllang}[2][0.5\textwidth]%
              {\hfill\hspace*{1.5em}%
               \begin{minipage}[t]{#1}{\small%
                          \begin{list}{(}{\ListNullAbstaende%
                                          \settowidth{\leftmargin}{(}%
                                          \settowidth{\labelwidth}{(}%
                                          \settowidth{\labelsep}{}%
                                         }%
                          \item#2)%
                          \end{list}}%
               \end{minipage}\\[-0.9ex]
              }%
\newcommand{\DefBemUmgeb}[1]%
           {\newenvironment{#1}[1][]%
                           {\begin{Einrueckung}{{\bf #1}}%
                            \ifx##1\empty\else{{\bf ##1}
                            
                                                        }\fi%
                            }%
                           {\end{Einrueckung}}}
\newcommand{\DefSBemUmgeb}[2]
           {\newenvironment{#1}[1][]%
                           {\begin{Einrueckung}{{\bf #2}}%
                            \ifx##1\empty\else{{\bf ##1}
                            
                                                        }\fi%
                            }%
                           {\end{Einrueckung}}}
\makeatletter
\newcommand{\DefBspUmgeb}[3]
           {\newcounter{#2}[#3]%
            \newenvironment{#1}[1][]%
                           {\stepcounter{#2}%
                            \renewcommand{\ZaehlerMarke}{\arabic{#2}}%
                            \renewcommand{\Einzugsname}{{\bf #1 \ZaehlerMarke}}%
                            \begin{Einrueckung}{\Einzugsname}
                            \ifx##1\empty\else{{\bf ##1}\\}\fi%
                            \renewcommand{\@currentlabel}{\ZaehlerMarke}%
                            }%
                           {\end{Einrueckung}}}
\makeatother
\newcommand{\ZaehlerbisEbene}{section}
\newcommand{\Ebenea}{section}
\newcommand{\Ebeneb}{subsection}

\newcommand{\Abschnittnummer}{%
            \ifx\ZaehlerbisEbene\Ebenea{\arabic{section}}%
             \else{%
              \ifx\ZaehlerbisEbene\Ebeneb{\arabic{section}.\arabic{subsection}}%
               \else{\arabic{section}.\arabic{subsection}.\arabic{subsubsection}}%
              \fi}%
            \fi}     
\newcommand{\Abschnittnummerpunkt}{\Abschnittnummer.}     
\newcommand{\Einzugsname}{}
\newcommand{\ZaehlerMarke}{}
\makeatletter
\newcommand{\DefThmUmgeb}[3]%
           {\newcounter{#1}[#3]%
            \newenvironment{#1}[1][]%
                           {\stepcounter{#2}%
                            \setcounter{#1}{\value{#2}}%
                            \renewcommand{\ZaehlerMarke}{\Abschnittnummerpunkt\arabic{#1}}%
                            \renewcommand{\Einzugsname}{{\bf #1 \ZaehlerMarke}}%
                            \begin{Einrueckung}{\Einzugsname}
                            \ifx##1\empty\else{{\bf ##1}
                            
                                                        }\fi%
                            \renewcommand{\@currentlabel}{\ZaehlerMarke}%
                            }%
                           {\end{Einrueckung}}}
\makeatother
\makeatletter
\newcommand{\DefSThmUmgeb}[4]%
           {\newcounter{#1}[#3]%
            \newenvironment{#1}[1][]%
                           {\stepcounter{#2}%
                            \setcounter{#1}{\value{#2}}%
                            \renewcommand{\ZaehlerMarke}{\Abschnittnummerpunkt\arabic{#1}}%
                            \renewcommand{\Einzugsname}{{\bf #4 \ZaehlerMarke}}
                            \begin{Einrueckung}{\Einzugsname}
                            \ifx##1\empty\else{{\bf ##1}

                                                        }\fi%
                            \renewcommand{\@currentlabel}{\ZaehlerMarke}%
                            }%
                           {\end{Einrueckung}}}
\makeatother
\makeatletter
\newcommand{\DefUnterNumThmUmgeb}[5]%
           {\newcounter{#1}[#3]%
            \newcounter{#4}%
            \newenvironment{#1}[1][]%
                           {\ifx##1\empty\else{\stepcounter{#2}\setcounter{#4}{0}}\fi%
                            \stepcounter{#4}%
                            \setcounter{#1}{\value{#2}}%
                            \renewcommand{\ZaehlerMarke}{\Abschnittnummerpunkt\arabic{#1}\alph{#4}}%
                            \renewcommand{\Einzugsname}{{\bf #5 \ZaehlerMarke}}
                            \begin{Einrueckung}{\Einzugsname}
                            \renewcommand{\@currentlabel}{\ZaehlerMarke}%
                            }%
                           {\end{Einrueckung}}}
\makeatother
\newenvironment{Beweis}[1][]%
               {\begin{Einrueckung}{{\bf Beweis}}%
                \ifx#1\empty\else{{\bf #1}

                                            }\fi%
                }%
               {\end{Einrueckung}%
                }%
\newenvironment{Proof}[1][]%
               {\begin{Einrueckung}{{\bf Proof}}%
                \ifx#1\empty\else{{\bf #1}

                                            }\fi%
                }%
               {\end{Einrueckung}%
                }%
               {\begin{Einrueckung}{{\bf \glqq Beweis\grqq}}%
                \ifx#1\empty\else{{\bf #1}
                
                                            }\fi%
                }%
               {\end{Einrueckung}%
                }%
               {\begin{Einrueckung}{{\bf Begr"undung}}%
                \ifx#1\empty\else{{\bf #1}
                
                                            }\fi%
                }%
               {\end{Einrueckung}%
                }%
\newenvironment{Hinrichtung}%
               {\begin{Einrueckungpur}{$\impliz$}}%
               {\end{Einrueckungpur}}%
\newenvironment{Rueckrichtung}%
               {\begin{Einrueckungpur}{$\invimpliz$}}%
               {\end{Einrueckungpur}}%
               {\begin{Einrueckungpur}{\glqq$\teilmenge$\grqq}}%
               {\end{Einrueckungpur}}%
               {\begin{Einrueckungpur}{\glqq$\obermenge$\grqq}}%
               {\end{Einrueckungpur}}%
               {\begin{Einrueckungpur}{"$\teilmenge$"}}%
               {\end{Einrueckungpur}}%
               {\begin{Einrueckungpur}{"$\obermenge$"}}%
               {\end{Einrueckungpur}}%
\newcommand{\qed}{\nopagebreak\hspace*{2em}\hspace*{\fill}{\bf qed}}
\newcommand{\ARabic}{\arabic}
\newcommand{\Nummerntypa}{\arabic}   
\newcommand{\Nummerntypb}{\alph}
\newcommand{\Nummerntypc}{\roman}
\newcommand{\Nummerntypd}{\Alph}

\newcommand{\Nra}{\Nummerntypa{Nummera}}            %
\newcommand{\Nrb}{\Nummerntypb{Nummerb}}            %
\newcommand{\Nrc}{\Nummerntypc{Nummerc}}                
\newcommand{\Nrd}{\Nummerntypd{Nummerd}}                
\newcommand{\ZeichenzuNrTyp}[1]%
           {\ifx#1\ARabic {.}\else{)}%
                  \fi}                              %
\newcommand{\NrZeicha}{\ZeichenzuNrTyp{\Nummerntypa}}
\newcommand{\NrZeichb}{\ZeichenzuNrTyp{\Nummerntypb}}
\newcommand{\NrZeichc}{\ZeichenzuNrTyp{\Nummerntypc}}
\newcommand{\NrZeichd}{\ZeichenzuNrTyp{\Nummerntypd}}
\newcommand{\ListMarkea}%
           {\Nra\NrZeicha}
\newcommand{\ListMarkeb}%
           {\Nra\NrZeicha\Nrb\NrZeichb}
\newcommand{\ListMarkec}%
           {\Nra\NrZeicha\Nrb\NrZeichb\Nrc\NrZeichc}
\newcommand{\ListMarked}%
           {\Nra\NrZeicha\Nrb\NrZeichb\Nrc\NrZeichc\Nrd\NrZeichd}
\newcommand{\Anfangszeichen}{}
\newcommand{\Anfangspunkt}{}
\newcounter{Schachtelebene}
\newcounter{Hilfszaehler}
\newcommand{\Hilfsbefehl}{}
\newcommand{\Schachtelebene}{\alph{Schachtelebene}}
\makeatletter
\newenvironment{AllgNumerierteListe}[2][]
               {\addtocounter{Schachtelebene}{1}%
		\setcounter{Hilfszaehler}{#2}%
                \renewcommand{\Anfangszeichen}%
                             {\renewcommand{\Hilfsbefehl}{\csname Nummerntyp\Schachtelebene \endcsname}%
                              \Hilfsbefehl{Hilfszaehler}}%
                \renewcommand{\Anfangspunkt}%
                             {\csname NrZeich\Schachtelebene \endcsname}%
                \begin{list}%
                      {\stepcounter{Nummer\Schachtelebene}%
                       \csname Nr\Schachtelebene \endcsname
                       \csname NrZeich\Schachtelebene \endcsname
                       }%
                      {\settowidth{\leftmargin}{M\Anfangszeichen\Anfangspunkt}%
                       \settowidth{\labelwidth}{\Anfangszeichen\Anfangspunkt}%
                       \settowidth{\labelsep}{M}%
                       \setlength{\topsep}{0pt}%
                       \setlength{\parskip}{0pt}%
                       \setlength{\partopsep}{0pt}%
                       \setlength{\itemsep}{0pt}%
                       \setlength{\parsep}{0pt}%
                      }%
                \renewcommand{\@currentlabel}{\csname ListMarke\Schachtelebene \endcsname}%
                }%
               {\ifthenelse{\equal{}{}}{\setcounter{Nummer\Schachtelebene}{0}}{}
                \addtocounter{Schachtelebene}{-1}%
                \end{list}}
\makeatother
\newenvironment{NumerierteListe}[1]
               {\begin{AllgNumerierteListe}{#1}}
               {\end{AllgNumerierteListe}}
\newenvironment{WeiterNumerierteListe}[1]
               {\begin{AllgNumerierteListe}[Weiter]{#1}}
               {\end{AllgNumerierteListe}}

\newcommand{\UnnumAnfangszeichen}{}
\newcounter{UnnumSchachtelebene}
\newcommand{\UnnumSchachtelebene}{\alph{UnnumSchachtelebene}}
\makeatletter
\newenvironment{UnnumerierteListe}%
               {\addtocounter{UnnumSchachtelebene}{1}%
                \renewcommand{\UnnumAnfangszeichen}%
                             {\csname UnnumZeich\UnnumSchachtelebene \endcsname}%
                \begin{list}%
                      {\UnnumAnfangszeichen}%
                      {\settowidth{\leftmargin}{M\UnnumAnfangszeichen}%
                       \settowidth{\labelwidth}{\UnnumAnfangszeichen}%
                       \settowidth{\labelsep}{M}%
                       \setlength{\topsep}{0pt}%
                       \setlength{\parskip}{0pt}%
                       \setlength{\partopsep}{0pt}%
                       \setlength{\itemsep}{0pt}%
                       \setlength{\parsep}{0pt}%
                      }%
                }%
               {\addtocounter{UnnumSchachtelebene}{-1}%
                \end{list}}
\makeatother
\newlength{\fktdefhilfslaenge}
\newcommand{\ohnefktdef}[4]
           {\hspace*{\fill}
            $\begin{array}[t]{ccc}%
            #1 & \nach & #2 \\
            #3 & \auf  & #4
            \end{array}$
            \hspace*{\fill}}
\newcommand{\fktdef}[5]
           {\hspace*{\fill}
            $\begin{array}[t]{cccc}%
            #1: & #2 & \nach & #3 \\    
                & #4 & \auf  & #5
            \end{array}$
            \settowidth{\fktdefhilfslaenge}{$#1$:}
            \hspace*{0.6 \fktdefhilfslaenge}  
            \hspace*{\fill}}
\newcommand{\fktdefpur}[5]
           {$\begin{array}[t]{cccc}%
            #1: & #2 & \nach & #3 \\    
                & #4 & \auf  & #5
            \end{array}$}
\newcommand{\fktdefabgesetztpur}[5]
           {
            
            $\begin{array}[t]{cccc}%
            #1: & #2 & \nach & #3 \\    
                & #4 & \auf  & #5
            \end{array}$
            \settowidth{\fktdefhilfslaenge}{$#1$:}
            \hspace*{0.6 \fktdefhilfslaenge}
            
           }
\newcommand{\fktdefabgesetzt}[5]
           {
           
            \hspace*{\fill}
            $\begin{array}[t]{cccc}%
            #1: & #2 & \nach & #3 \\    
                & #4 & \auf  & #5
            \end{array}$
            \settowidth{\fktdefhilfslaenge}{$#1$:}
            \hspace*{0.6 \fktdefhilfslaenge}  
            \hspace*{\fill}
            
            }
\newcommand{\ohnefktdefabgesetzt}[4]
           {      

            \hspace*{\fill}
            $\begin{array}[t]{ccc}%
            #1 & \nach & #2 \\
            #3 & \auf  & #4
            \end{array}$
            \hspace*{\fill}

            }
\newcommand{\doppelohnefktdefabgesetzt}[6]
           {

            \hspace*{\fill}
            $\begin{array}[t]{ccccc}%
            #1 & \nach & #2 & \nach & #3\\
            #4 & \auf  & #5 & \auf  & #6
            \end{array}$
            \hspace*{\fill}

            }
\newcommand{\anhang}%
           {\appendix
            \sectioninh{Anhang}
            \renewcommand{\Abschnittnummer}{%
                  \ifx\ZaehlerbisEbene\Ebenea{\Alph{section}}%
                  \else{%
                        \ifx\ZaehlerbisEbene\Ebeneb{\Alph{section}.\arabic{subsection}}%
                        \else{\Alph{section}.\arabic{subsection}.\arabic{subsubsection}}%
                        \fi}%
                  \fi}%
            \renewcommand{\Abschnittnummerpunkt}{\Abschnittnummer.}     
            }            
\newcommand{\anhangengl}%
           {\appendix
            \sectioninh{Appendix}
            \renewcommand{\Abschnittnummer}{%
                  \ifx\ZaehlerbisEbene\Ebenea{\Alph{section}}%
                  \else{%
                        \ifx\ZaehlerbisEbene\Ebeneb{\Alph{section}.\arabic{subsection}}%
                        \else{\Alph{section}.\arabic{subsection}.\arabic{subsubsection}}%
                        \fi}%
                  \fi}%
            \renewcommand{\Abschnittnummerpunkt}{\Abschnittnummer.}     
            }

\newcounter{wdhlstufe}
\newcommand{\sectioninh}[1]%
           {\section*{#1}%
            \addcontentsline{toc}{section}{#1}}
\newcommand{\bezeichnung}[3]%
           {\begin{Einrueckungpur}{\hbox to 6em{#1}\hbox to 2.4em{\hfill#2}}
            #3
            \end{Einrueckungpur}}

\newcommand{\doppelteinfach}{e}

\newcommand{\ifdoppelt}[1]{\ifthenelse{\equal{\doppelteinfach}{d}}{#1}{}}
\newcommand{\ifeinfach}[1]{\ifthenelse{\equal{\doppelteinfach}{e}}{#1}{}}

\newlength{\querfhilfsl}              %

\newlength{\hll}

\DefThmUmgeb{Theorem}{Theorem}{\ZaehlerbisEbene}
\DefThmUmgeb{Definition}{Definition}{\ZaehlerbisEbene}
\DefThmUmgeb{Satz}{Theorem}{\ZaehlerbisEbene}
\DefThmUmgeb{Proposition}{Theorem}{\ZaehlerbisEbene}
\DefThmUmgeb{Lemma}{Theorem}{\ZaehlerbisEbene}
\DefThmUmgeb{Folgerung}{Theorem}{\ZaehlerbisEbene}
\DefThmUmgeb{Corollary}{Theorem}{\ZaehlerbisEbene}
\DefThmUmgeb{Vorschrift}{Definition}{\ZaehlerbisEbene}
\DefThmUmgeb{Construction}{Definition}{\ZaehlerbisEbene}
\DefSThmUmgeb{FormSatz}{Theorem}{\ZaehlerbisEbene}{\glqq Satz\grqq} 
\DefThmUmgeb{Vermutung}{Theorem}{\ZaehlerbisEbene}
\DefThmUmgeb{Conjecture}{Theorem}{\ZaehlerbisEbene}
\DefThmUmgeb{Konvention}{Definition}{\ZaehlerbisEbene}
\DefThmUmgeb{Feststellung}{Theorem}{\ZaehlerbisEbene}
\DefUnterNumThmUmgeb{DefinitionZusatzNum}{Definition}{\ZaehlerbisEbene}{DefZN}{Definition}
\DefBspUmgeb{Beispiel}{Beispiel}{subsubsection}
\DefBspUmgeb{Example}{Example}{subsubsection}
\DefBspUmgeb{Frage}{Frage}{section}
\DefBspUmgeb{Question}{Question}{section}
\DefBspUmgeb{Aufgabe}{Aufgabe}{section}
\DefBspUmgeb{Ziel}{Ziel}{section}
\DefBemUmgeb{Bemerkung}
\DefBemUmgeb{Remark}
\DefSBemUmgeb{OffeneFrage}{Offene Frage}
\newcommand{\bdf}{\begin{Definition}}
\newcommand{\edf}{\end{Definition}}
\newcommand{\bvorsch}{\begin{Vorschrift}}
\newcommand{\evorsch}{\end{Vorschrift}}
\newcommand{\bconst}{\begin{Construction}}
\newcommand{\econst}{\end{Construction}}
\newcommand{\bthm}{\begin{Theorem}}
\newcommand{\ethm}{\end{Theorem}}
\newcommand{\bsatz}{\begin{Satz}}
\newcommand{\esatz}{\end{Satz}}
\newcommand{\bprop}{\begin{Proposition}}
\newcommand{\eprop}{\end{Proposition}}
\newcommand{\blem}{\begin{Lemma}}
\newcommand{\elem}{\end{Lemma}}
\newcommand{\bfolg}{\begin{Folgerung}}
\newcommand{\efolg}{\end{Folgerung}}
\newcommand{\bcorr}{\begin{Corollary}}
\newcommand{\ecorr}{\end{Corollary}}
\newcommand{\bfest}{\begin{Feststellung}}
\newcommand{\efest}{\end{Feststellung}}
\newcommand{\bbew}{\begin{Beweis}}
\newcommand{\ebew}{\end{Beweis}}
\newcommand{\bpf}{\begin{Proof}}
\newcommand{\epf}{\end{Proof}}
\newcommand{\bwnum}{\begin{WeiterNumerierteListe}}
\newcommand{\ewnum}{\end{WeiterNumerierteListe}}
\newcommand{\bdfzn}{\begin{DefinitionZusatzNum}}
\newcommand{\edfzn}{\end{DefinitionZusatzNum}}
\newcommand{\bbem}{\begin{Bemerkung}}
\newcommand{\ebem}{\end{Bemerkung}}
\newcommand{\brem}{\begin{Remark}}
\newcommand{\erem}{\end{Remark}}
\newcommand{\bnum}{\begin{NumerierteListe}}
\newcommand{\enum}{\end{NumerierteListe}}
\newcommand{\bunum}{\begin{UnnumerierteListe}}
\newcommand{\eunum}{\end{UnnumerierteListe}}
\newcommand{\bbsp}{\begin{Beispiel}}
\newcommand{\ebsp}{\end{Beispiel}}
\newcommand{\bex}{\begin{Example}}
\newcommand{\eex}{\end{Example}}
\newcommand{\bfrag}{\begin{Frage}}
\newcommand{\efrag}{\end{Frage}}
\newcommand{\bquest}{\begin{Question}}
\newcommand{\equest}{\end{Question}}
\newcommand{\baufg}{\begin{Aufgabe}}
\newcommand{\eaufg}{\end{Aufgabe}}
\newcommand{\bof}{\begin{OffeneFrage}}
\newcommand{\eof}{\end{OffeneFrage}}
\newcommand{\bverm}{\begin{Vermutung}}
\newcommand{\everm}{\end{Vermutung}}
\newcommand{\bconj}{\begin{Conjecture}}
\newcommand{\econj}{\end{Conjecture}}
\newcommand{\bkonv}{\begin{Konvention}}
\newcommand{\ekonv}{\end{Konvention}}
\newcommand{\bglklein}{\begin{StandardUnnumGleichungKlein}}
\newcommand{\eglklein}{\end{StandardUnnumGleichungKlein}}
\newcommand{\bgl}{\begin{StandardUnnumGleichung}}
\newcommand{\egl}{\end{StandardUnnumGleichung}}
\newcommand{\bglrtext}{\begin{XrelYZNumGleichung}}
\newcommand{\eglrtext}{\end{XrelYZNumGleichung}}

\newcommand{\berlgl}{\begin{StandardUnnumGleichung}}
\newcommand{\eerlgl}{\end{StandardUnnumGleichung}}
\newcommand{\beinrueck}{\begin{Einrueckungpur}} 
\newcommand{\eeinrueck}{\end{Einrueckungpur}}
\newcommand{\beinflist}{\begin{EinfachListe}} 
\newcommand{\eeinflist}{\end{EinfachListe}}
\newcommand{\beq}{\begin{equation}}
\newcommand{\eeq}{\end{equation}}
\newcommand{\bhin}{\begin{Hinrichtung}}
\newcommand{\ehin}{\end{Hinrichtung}}
\newcommand{\brueck}{\begin{Rueckrichtung}}
\newcommand{\erueck}{\end{Rueckrichtung}}
\newcommand{\bvl}{\begin{AutoLabelLaengenListe}{\ListNullAbstaende}}
\newcommand{\evl}{\end{AutoLabelLaengenListe}}
\newcommand{\df}[1]{{\bf #1}}

\newlength{\adressabstand}
\newcommand{\minimallaenge}{L}
\newcommand{\wmers}{W}   
\newcommand{\esp}{I}   
\newcommand{\doppelbinom}{d}
\newcommand{\wechsel}{{\cal C}}
\newcommand{\kwechsel}{{\cal D}}
\newcommand{\Bignorm}[1]{\Big|\Big|#1\Big|\Big|}
\newcommand{\Bigbetrag}[1]{\Big|#1\Big|}
\begin{document}
\title{Asymptotic Positivity of Hurwitz Product Traces: Two Proofs}
\author{Christian Fleischhack${}^{1,\ast}$ and
        Shmuel Friedland${}^{2,3,}$%
	\thanks{e-mail: {\tt christian.fleischhack@math.uni-hamburg.de}, {\tt friedlan@uic.edu}} \\
        \\
        {\normalsize\em ${}^1$Department Mathematik${}^{\phantom1}$}\\[\adressabstand]
        {\normalsize\em Universit\"at Hamburg}\\[\adressabstand]
        {\normalsize\em Bundesstra\ss e 55}\\[\adressabstand]
        {\normalsize\em 20146 Hamburg, Germany}
        \\[-25\adressabstand]      
        {\normalsize\em ${}^2$Department of Mathematics, Statistics, and Computer Science${}^{\phantom2}$}\\[\adressabstand]
        {\normalsize\em University of Illinois at Chicago}\\[\adressabstand]
        {\normalsize\em Chicago, IL 60607-7045, USA}
        \\[-25\adressabstand]      
        {\normalsize\em ${}^3$Berlin Mathematical School${}^{\phantom3}$}\\[\adressabstand]
        {\normalsize\em Institut f\"{u}r Mathematik}\\[\adressabstand]
        {\normalsize\em Technische Universit\"{a}t Berlin}\\[\adressabstand]
        {\normalsize\em Stra\ss e des 17. Juni 136}\\[\adressabstand]
        {\normalsize\em 10623 Berlin, Germany}
        \\[-25\adressabstand]}

\date{October 31, 2008}
\maketitle

\newcommand{\bsmat}{\bigl(\begin{smallmatrix}}
\newcommand{\esmat}{\end{smallmatrix}\bigr)}
\newcommand{\falsch}{{\bf falsch!!!!!!!!}}
\newcommand{\neueseite}{\newpage}
\newcommand{\word}{{\cal W}}
\newcommand{\irgendeinefunktion}{{\xi}}
\newcommand{\trm}{\tr\,}

\begin{abstract}
Consider the polynomial $\trm (A + tB)^m$ in $t$ for 
positive hermitian matrices $A$ and $B$ with $m \in \N$.
The Bessis-Moussa-Villani conjecture (in the equivalent form 
of Lieb and Seiringer) states that this polynomial has
nonnegative coefficients only. 
We prove that they are at least asymptotically positive, for the
nontrivial case of $AB \neq \NULL$.
More precisely, 
we show 
---once complex-analytically, once combinatorially---
that the $k$-th coefficient is positive 
for all integer $m \geq m_0$, where $m_0$ depends on $A$, $B$ and $k$.
\end{abstract}

\vspace*{\fill}

\begin{center}
\begin{minipage}{0.59\textwidth}
\noindent
This article merges the following two original articles:
\bunum
\item
\cite{paper27} ({\tt arxiv:0804.3665}) by Christian Fleischhack,
\item
\cite{bmv25} ({\tt arxiv:0804.3948}) by Shmuel Friedland.
\eunum
\end{minipage}
\end{center}

\newpage


\section{Introduction}

Some 30 years ago, 
Bessis, Moussa and Villani (BMV) conjectured \cite{bmv1}%
\footnote{Originally, in \cite{bmv1}, a stronger conjecture has been
stated: For any bounded-from-below self-adjoint operators $A$ and
$B$ and any eigenvector $\varphi$ of $B$, the function
$\skalprod{\varphi}{\e^{-(A + t B)} \varphi}$ is 
the Laplace transform of a positive measure $\mu$ whose support 
is contained in the convex hull of the spectrum of $B$. This
conjecture, however, turned out to be wrong as seen by Froissart
(see the notes added in proof in \cite{bmv1}; alternatively,
see the example given in \cite{bmv12}). 
Then, BMV conjectured that, nevertheless, the statement 
remains valid for the trace.}
that 
for any hermitian $n \kreuz n$ matrices $A$ and $B$,
the function
\bgl
\mu(t) & := & \trm \exp(A - t B)
\egl\noindent
with $t \in \R$ 
is the Laplace transform of a positive measure on $[0,\infty)$, provided
$B$ is positive%
\footnote{Positivity of a hermitian matrix 
$B$ means that $\skalprod{x}{Bx} \geq 0$ for all
$x\in\C^n$. In particular, $\NULL$ is positive.}.
Lieb and Seiringer \cite{bmv8}
proved that this statement is equivalent to the assertion that,
for positive integers $m$ and positive
hermitian $A$ and $B$,
the polynomial 
\bglklein
\trm(A + tB)^m & = & \sum_k \trm S_{m,k}(A,B) \: t^k
\eglklein\noindent
has nonnegative coefficients only. Here, the
Hurwitz product $S_{m,k}$ \cite{bmv16} equals the sum of all 
words in $A$ and $B$, containing $m-k$ letters $A$ and $k$ letters $B$.
Although the conjecture is widely expected to be true, 
there are, by now, only partial results confirming it. 
Of course, it is true in the obvious cases 
of commuting $A$ and $B$, for
$n=1$ and for $k \leq 2$. For $n = 2$, the statement follows since
there is a common basis 
where $A$ and $B$ have nonnegative entries only \cite{bmv8}.
Beyond that, positive results have been obtained for lower $m$; at present,
the conjecture is proven for $m \leq 13$ \cite{bmv20,bmv16}. 
This relied on 
two main ideas: First, generally, if the conjecture is given for some $(m,k)$,
then it holds for any $(m',k')$ with $m' \leq m$, $k' \leq k$ and 
$m'-k' \leq m-k$ \cite{bmv16}. Second,
more specifically, H\"agele \cite{bmv17} proposed to 
write $S_{m,k}(A,B)$ ---up to some cyclic permutations--- as a sum of 
positive terms.
Although not possible for $(6,3)$ and several other cases \cite{bmv21},
he was able to find such a decomposition 
for $(7,3)$, implying the BMV conjecture for $m\leq 7$.
More refined methods \cite{bmv20} using computer algebra 
established the cases $(14,4)$ and $(14,6)$, 
implying the conjecture for $m \leq 13$. Recently, it has been shown that 
the conjecture is always true for $k = 4$ \cite{bmv22}, implying it for $k = 3$.
Other results show that
one may restrict oneself to the case of singular matrices $A$ and $B$
when proving the conjecture inductively \cite{bmv16}.
Although the BMV conjecture is still open, it is known that
the untraced coefficients $S_{m,k}(A,B)$ need not be positive.
The easiest example is $S_{6,3}(A,B)$ 
for appropriate $A$ and $B$; 
here,
some single
words may even have negative trace \cite{bmv6}.

In the present paper we study a different side of the problem. 
Shifting the focus from (computer) algebra back to analysis, we 
are going to investigate the behaviour of the terms $\trm S_{m,k}(A,B)$
for large 
instead of small $m$. 
Our main result is%
\footnote{If the dimensions of the matrices $\NULL$ and $\EINS$ 
should be clear from the context, we may refrain from specifying them
by writing $\NULL_n$ and $\EINS_n$, respectively.}

\bthm
\label{thm:positiv_klein}
Let $A$ and $B$ be positive hermitian $n \kreuz n$ matrices and $k \in \N$.\

Then there is some $m_0 \in \N$, such that:

\bgl[0.5ex]
AB = \NULL & \impliz & \trm S_{m,k}(A,B) = 0 \: \text{ for any integer $m \neq k \neq 0$. } \s
AB \neq \NULL & \impliz & \trm S_{m,k}(A,B) > 0 \: \text{ for any integer $m \geq m_0$. }
\egl
\ethm

\noindent
We are going to prove this theorem in two different ways --- once using complex-analysis methods,
once using combinatorics. In the latter case we also give a concrete 
estimate for $m_0$.
Let us summarize the main ideas of the proofs.
Since the case $AB = \NULL$ is trivial, we may assume $AB \neq \NULL$.
Moreover, we may assume that $A$ has unit norm.%
\footnote{In the main text, we always consider the operator norm. 
Other norms will be discussed in the context of Euler-Lagrange equations
in the appendix.}

For the combinatorial proof observe that,
if $m$ increases,
the $k$ letters $B$ are getting more and more sparsely distributed inside the
words in $S_{m,k}(A,B)$. Indeed, most of the terms are of the form
$A^{i_1} B A^{i_2} \cdots B A^{i_{k+1}}$ with rather large $i_\iota$.
These words are approximated by the positive hermitian matrix $(P_A B P_A)^k$, where $P_A$ is the
hermitian projector $\lim_{i \gegen \infty} A^i$. The assertion follows
unless $\trm (P_A B)^k = \trm (P_A B P_A)^k \geq 0$ vanishes. But, then 
$A = \EINS_{n-l} \dirsum A'$ and $B = \NULL_{n-l} \dirsum B'$
for some 
positive hermitian $l \kreuz l$ matrices 
$A'$, $B'$ with $0 < l < n$, such that the proof follows inductively.

For the complex-analytic proof, again by induction, 
we also may assume $\trm (P_A B)^k > 0$.
Consider the series
\bgl
\sum_{\laufm=k}^{\infty} \: \frac{k}{\laufm}\: \trm S_{\laufm,k}(A,B) \:\: \tau^{\laufm-k} 
 & = & \trm\bigl[B (\EINS - \tau A)^{-1}\bigr]^k 
 \breitrel= \frac{\trm (P_A B)^k}{(1 - \tau)^k}  
          + \sum_{\laufj=0}^{k-1} \frac{\rest_\laufj(\tau)}{(1 - \tau)^\laufj}\,.
\egl\noindent
Here, 
each $\rest_\laufj$ is a rational function vanishing at infinity and
having poles only outside the closed unit disk.
Now the proof follows, since
the $\laufm$-th Taylor coefficient of $(1 - \tau)^{-\laufj}$ is a polynomial in $\laufm$ 
of degree $\laufj - 1$ with positive leading coefficient.

Unfortunately, the dependence of $m_0$ on $A$ and $B$ is crucial for our
proofs of the theorem. Therefore, the full BMV conjecture does not follow directly 
from the theorem above. 
Nevertheless, some (admittedly, simple) numerical
simulations indicate further structures in the sequence of 
$\trm S_{m,k}(A,B)$ for general $k$. 
To see them, we should first factor out the
trivial dependencies. In fact, observe that otherwise this term
(in general) diverges; we have, e.g.,
$\trm S_{m,k}(\kappa \EINS,\lambda \EINS) 
    = n \kappa^{m-k} \lambda^k \binom m k$.
Thus, we will study the normalized quotient
\bgl
q_{m,k}(A,B) & := & \frac{\trm S_{m,k}(A,B)}{\trm S_{m,k}(\norm{A} \EINS,\norm{B} \EINS)} \:,
\egl\noindent
as the BMV conjecture is now equivalent to $q_{m,k}(A,B) \geq 0$ for
all positive hermitian matrices $A$ and $B$ having norm $1$.
Since the theorem above tells us that $q_{m,k}(A,B) > 0$
for sufficiently large $m$, the BMV conjecture would now follow 
if one could establish

\bconj
Let $A$ and $B$ be positive hermitian $n \kreuz n$ matrices with $AB \neq \NULL$.

Then, for any fixed $k\in\N$, the sequence 
\bgl
\bigl(q_{m+k,k}(A,B)\bigr){}_{m\in\N}
\egl
is 
decreasing.
\econj

Despite to the mentioned 
numerical hints, we have not been able to prove this conjecture analytically.
Nevertheless, 
we have been able to deduce further properties of
$q_{m,k}(A,B)$ for large $m$ and general $k$:

\bthm
\label{thm:asymptotik}
Let $\varepsilon > 0$, let $A$ and $B$ be nonzero positive hermitian
matrices, and let 
$d$ be the dimension of the intersection of the 
eigenspaces of $A$ and $B$ w.r.t.\ their highest eigenvalues. Then 
there are $m_0, k_0 >0$, such that
\bgl
q_{m,k}(A,B) & > & \frac dn -\varepsilon \breitrel{ \: } \text{ for $m \geq m_0$ and $k, m-k \geq 0$} \phantom{._0}
\egl
and 
\bgl
q_{m,k}(A,B) & < & \frac {d}n + \varepsilon \breitrel{ \: } \text{ for $m \geq m_0$ and $k, m-k \geq k_0$.}
\egl
\ethm
In particular, 
$\trm S_{m,k}(A,B)$ is strictly positive
for all $k$, provided $m$ is sufficiently large and the matrices 
$A$ and $B$ share a
common eigenvector w.r.t.\ the respective maximal nonzero eigenvalue.

Let us now sketch the idea of the proof for normalized $A$ and $B$.
If $d > 0$,
we may decompose $A$ and $B$ 
into $A' \dirsum \EINS_d$ and $B' \dirsum \EINS_d$,
respectively, where $A'$ and $B'$ are positive hermitian with $\norm{A'B'} < 1$.
Since
\bglklein
\trm S_{m,k}(A,B) 
 & = & \trm S_{m,k}(A',B') + \trm S_{m,k}(\EINS_d,\EINS_d) \\
 & = & \trm S_{m,k}(A',B') + \frac dn \: \trm S_{m,k}(\EINS_n,\EINS_n),
\eglklein\noindent
we may assume $d = 0$, i.e., $\norm{AB} < 1$.
Moreover, by Theorem \ref{thm:positiv_klein}, we may assume that
$k$ and $m-k$ are not too small.
Now, the typical element among the $S_{m,k}(A,B)$ terms contains a
higher and higher number of subwords $AB$. 
The norm estimate $\norm{AB}<1$ implies that 
$q_{m,k}(A,B)$, hence, the average contribution 
of a word to $S_{m,k}(A,B)$ is getting arbitrarily small.

\leerezeile

\noindent
Our paper is organized as follows:
First, for completeness, we collect some simple properties
of normalized positive hermitian matrices. 
Next, we study properties of certain power series whose coefficients are 
Hurwitz products or their traces.
Then we use combinatorial
methods to the calculate the number of words in $A$ and $B$ containing the
subword $AB$ a certain number of times, and derive estimates for these
figures. In Section \ref{sect:main_thms}, 
we prove the theorems announced above.
Finally, in Appendix \ref{app:eleq}, we derive the
Euler-Lagrange equations in a slightly more
abstract way than in \cite{bmv16} and extend these results to
several norms.


\section{Some Algebra}

In this section we review the asymptotic 
behaviour of powers of positive hermitian
matrices as well as of their products. Most importantly, we
will recall that $A^i$ for positive unit-norm matrices $A$ always tends
to the projector%
\footnote{Throughout the whole paper, any projector
is assumed to be a hermitian projector.}
onto the highest eigenspace (i.e., for the eigenvalue $1$); 
powers of matrix products converge to projectors to 
common highest eigenspaces. Moreover, we derive some 
norm and trace estimates as well as some criteria for the
product of two matrices to vanish.


\subsection{Power Limits}

\bdf
For any $n\kreuz n$ matrix $A$, define
$P_A := \lim_{i \gegen \infty} A^i$, if the limit exists.
\edf
Obviously, we have $P_\EINS = \EINS$ and $P_\NULL = \NULL$.


\blem
\label{lem:ew1}
Let $A$ be any $n \kreuz n$ matrix, such that $P_A$ exists.
Then we have:
\bunum
\item
$P_A$ is idempotent;
\item
$A P_A = P_A = P_A A$;
\item
$P_A x = x$ $\aequ$ $A x = x$, where $x\in\C^n$;
\item
$P_{U^{-1} A U}$ exists for any invertible
$n \kreuz n$ matrix $U$, and it equals $U^{-1} P_A U$.
\eunum
\elem
The final statement implies that we often may restrict ourselves to
the case of diagonal $A$, as long as we investigate $P_A$ for hermitian 
$A$.

\bpf
We have
\bglklein
P_A P_A 
 & = & \bigl(\lim_{i \gegen \infty} A^i\bigr) \bigl(\lim_{j \gegen \infty} A^j\bigr)
 \breitrel= \lim_{i \gegen \infty} \bigl(A^i \bigl(\lim_{j \gegen \infty} A^j\bigr)\bigr)\\
 & = & \lim_{i \gegen \infty} \bigl(\lim_{j \gegen \infty} A^{i+j}\bigr) 
 \breitrel= \lim_{i \gegen \infty} P_A
 \breitrel= P_A
\eglklein
and, similarly, $A P_A = P_A = P_A A$. Now,
$x = P_A x$ implies 
\bgl
A x \breitrel= A (P_A x) 
    \breitrel= (A P_A) x 
    \breitrel= P_A x
    \breitrel= x    .
\egl
The remaining assertions are obvious.
\qed
\epf

\bdf
Let $A$ be any $n\kreuz n$ matrix.

Then $\esp_\lambda(A)$ denotes its eigenspace in $\C^n$ for the eigenvalue
$\lambda$.
\edf

\blem
\label{lem:P_A_ex_fuer_hermit_leq_1}
If $A$ is hermitian with $\norm{A} \leq 1$ and if $-1$ is not
in the spectrum of $A$, 
then $P_A$ exists and is a projector.
Moreover, $\im P_A = \esp_1(A)$. 
\elem
\bpf
Consider $A$ in diagonal form and use Lemma \ref{lem:ew1}.
\qed
\epf

\blem
\label{lem:proj_dirsum}
Let $A_1$ be an $n_1\kreuz n_1$ matrix and $A_2$ be an $n_2\kreuz n_2$ matrix,
such that $P_{A_1}$ and $P_{A_2}$ exist.
Then $P_{A_1 \dirsum A_2}$ exists and equals $P_{A_1} \dirsum P_{A_2}$.
\elem
\bpf
Use that $(A_1 \dirsum A_2)^i = A_1^i \dirsum A_2^i$.
\qed
\epf


\subsection{Phone Matrices}

\bdf
A matrix $A$ is called \df{$n$-phone} iff $A$ is a positive, hermitian 
$n\kreuz n$ matrix whose largest eigenvalue is $1$.
\edf
Recall that the norm of a positive hermitian matrix coincides with its
largest eigenvalue.
\blem
Any nonzero 
projector is an $n$-phone matrix.
\elem

\blem
\label{lem:potenz_diff_n-phone_proj}
For any $k \in \N_+$ and any 
$n$-phone matrix $A$, we have 
\bgl
(A - P_A)^k \breitrel= A^k - P_A
 & \text{ and } & \norm{A^k - P_A} \breitrel= \norm{A - P_A}^k
\egl
\elem
\bpf
This follows inductively, using
\bgl
(A - P_A)^{k + 1} 
  & = & (A - P_A)^k (A - P_A) 
  \breitrel= (A^k - P_A)(A - P_A) \\
  & = & A^{k+1} - P_A A - A^k P_A + P_A P_A 
  \breitrel= A^{k+1} - P_A 
\egl
by Lemma \ref{lem:ew1}.
The norm equality now follows from $\norm{M^k} = {\norm M}^k$ for any hermitian matrix $M$.
\qed
\epf
Occasionally, we will decompose matrices into direct sums of matrices.
When we simply state that some matrix $B$ equals $B_1 \dirsum B_2$,
then we tacitly assume that there is some decomposition of $\C^n$ into
$X_1 \dirsum X_2 \iso \C^{\dim X_1} \dirsum \C^{\dim X_2}$,
such that $B \einschr{X_i} = B_i : X_i \nach X_i$.
Furthermore, note that whenever we decompose several matrices into direct sums,
we will always assume that all these matrices are decomposed w.r.t.\ 
one and the same decomposition of $\C^n$.

\blem
\label{lem:decomp_phone_matrix}
Let $A$ be an $n$-phone matrix and let $P_A = \EINS_{n-l} \dirsum \NULL_{l}$
for some $0 \leq l \leq n$.

Then there is some $0 \leq \alpha < 1$ and some $l$-phone matrix $A'$,
such that $A$ equals $\EINS_{n-l} \dirsum \alpha A'$. Moreover, we have 
$l > 0$, unless $A = \EINS_n$, and $l < n$.
\elem

\bpf
\bunum
\item
If $l = n$, then $P_A = \NULL_n$, whence $\esp_1(A) = 0$ 
by Lemma \ref{lem:P_A_ex_fuer_hermit_leq_1}, i.e.,
$\norm A < 1$. 
\item
If $l = 0$, we have $P_A = \EINS_n$, i.e.,
$\esp_1(A) = \esp_1(P_A) = n$ and, therefore, $A = \EINS_n$.
\item
If $0 < l < n$, then 
$P_A = \bsmat \EINS & \NULL \\ \NULL & \NULL \esmat$. Let
$A = \bsmat F & G^\ast \\ G & H \esmat$ with positive hermitian 
matrices $F$ (of size $n-l$) and $H$ (of size $l$).
From $P_A A = P_A$, we derive $F = \EINS$ and $G = \NULL$,
whence $A = \EINS_{n-l} \dirsum H$. By 
\bgl
\EINS_{n-l} \dirsum \NULL_l 
 \breitrel= P_A 
 \breitrel= P_{\EINS_{n-l} \dirsum H} 
 \breitrel= P_{\EINS_{n-l}} \dirsum P_H 
 \breitrel= \EINS_{n-l} \dirsum P_H, 
\egl
we have $P_H = \NULL_l$, whence
$\norm H < 1$, again by Lemma \ref{lem:P_A_ex_fuer_hermit_leq_1}.
Now, define 
$\alpha := \norm H$ and $A' := \alpha^{-1} H$ (or
$A' = \EINS_l$ if $H = \NULL_l$).
\qed
\eunum
\epf

\bcorr
\label{corr:norm(a-pa)_kleiner_1}
For any $n$-phone matrix $A$, we have $\norm{A - P_A} < 1$.
\ecorr


\subsubsection{Shared Eigenspaces}

\blem
\label{lem:norm-ew1}
Let $A_1, \ldots, A_N$
be $n$-phone matrices and let $x\in\C^n$.
Then 
\bgl
\norm{A_N \cdots A_1 x} = \norm{x}
 & \aequ & A_i x = x \text{ for all $i=1,\ldots, N$.}
\egl
\elem
\bpf
We may assume that $x \neq 0$. Moreover, the $\invimpliz$ direction
is trivial.
We now prove the $\impliz$ statement by induction. 
Let $N = 1$ and denote shortly $A := A_1$.
Then there is a unitary
$U$, such that $D := U A U^\ast$ is diagonal. Setting $y := Ux$, we have
\bgl
\norm{Dy} 
 \breitrel\ident \norm{UAU^\ast Ux}
 \breitrel= \norm{Ax}
 \breitrel= \norm{x}
 \breitrel= \norm{Ux}
 \breitrel\ident \norm{y}.
\egl
Writing $D =: \diag(d_1,\ldots,d_n)$ with $0 \leq d_j \leq 1$ and $y =: (y_1,\ldots,y_n)^T$,
we find that $Dy = (d_1 y_1, \ldots, d_n y_n)^T$,
whence 
\bglklein
\sum_{j=1}^n (1 - d_j^2) \betrag {y_j}^2 
 & = & \sum_{j=1}^n \betrag {y_j}^2 - \sum_{j=1}^n d_j^2 \betrag {y_j}^2 
 \breitrel= \norm{y}^2 - \norm{Dy}^2
 \breitrel=0.
\eglklein
Since $0 \leq d_j \leq 1$, we have $(1 - d_j^2) \betrag {y_j}^2 = 0$ for all $j$.
Consequently,
\bgl
d_j = 1 \text{ or } y_j = 0 \text{ for all $j$}
& \impliz & (d_j - 1) y_j = 0 \text{ for all $j$}\\
& \impliz & d_j y_j = y_j \text{ for all $j$}\\
& \impliz & Dy = y.
\egl
Now, $Ax = U^\ast D U U^\ast y = U^\ast Dy = U^\ast y = x$.

Next, let $N>1$ and assume the assertion to be proven for $N-1$.
We now have 
\bgl
\norm{x} 
 & = &  \norm{A_N A_{N-1} \cdots A_1 x} 
 \breitrel\leq \norm{A_N} \norm{A_{N-1} \cdots A_1 x} \\
 & = & \norm{A_{N-1} \cdots A_1 x} 
 \breitrel\leq \norm{A_{N-1}} \cdots \norm{A_1} \norm{x} 
 \breitrel= \norm{x},
\egl
whence $\norm{A_{N-1} \cdots A_1 x} = \norm{x}$. By induction, $A_i x = x$
for all $i<N$. On the other hand, this implies 
$\norm{A_N x} = \norm{A_N A_{N-1} \cdots A_1 x} = \norm{x}$. From the
induction beginning, we get $A_N x = x$ as well.
\qed
\epf

\bcorr
\label{corr:phoneprod-eigenwerte}
For any $n$-phone matrices $A_1, \ldots, A_N$ we have
\bglklein
\esp_1(A_1 \cdots A_N) & = & \esp_1(A_1) \cap \ldots \cap \esp_1(A_N).
\eglklein
\ecorr
\bpf
\bunum
\item
Let $x \in \esp_1(A_1 \cdots A_N)$, i.e.\
$A_1 \cdots A_N x = x$. Lemma \ref{lem:norm-ew1} implies
$A_i x = x$ for all $i$.
\item
Trivial.
\qed
\eunum
\epf

\bcorr
\label{corr:phoneprod-norm1}
For any $n$-phone matrices $A_1, \ldots, A_N$ we have
\bglklein
\esp_1(A_1 \cdots A_N) \breitrel\neq 0
& \aequ & \norm{A_1 \cdots A_N} = 1.
\eglklein
\ecorr
\bpf
\bunum
\item
If $\norm{A_1 \cdots A_N} = 1$, then there is some nonzero $x\in\C^n$
with $\norm{A_1 \cdots A_N x} = \norm x$.
Lemma \ref{lem:norm-ew1} implies that $A_i x = x$ for all $i$.
This, of course, implies $A_1 \cdots A_N x = x$.
\item
If $\esp_1(A_1 \cdots A_N) \neq 0$, then, 
by Corollary \ref{corr:phoneprod-eigenwerte}, there is some nonzero $x\in\C^n$,
such that 
$A_i x = x$ for all $i$. Now, $A_1 \cdots A_N x = x$ 
and $\norm{A_1 \cdots A_N} = 1$.
\qed
\eunum
\epf

\bcorr
\label{corr:phoneprod-projector}
Let $A_1, \ldots, A_N$ be $n$-phone matrices. Then we have:
\bgl
\text{$P_{A_1 \cdots A_N}$ exists and equals $\NULL$.} 
  & \aequ & \esp_1(A_1 \cdots A_N) \breitrel=0.
\egl
\ecorr
\bpf
\bunum
\item
If $\esp_1(A_1 \cdots A_N) = 0$, then, 
by Corollary \ref{corr:phoneprod-norm1}, $\norm{A_1 \cdots A_N} < 1$,
whence we have $\norm{(A_1 \cdots A_N)^j} \leq \norm{A_1 \cdots A_N}^j \gegen 0$
for $j \gegen \infty$. Consequently, $P_{A_1 \cdots A_N} = \NULL$.
\item
If $\esp_1(A_1 \cdots A_N) \neq 0$, then,
by Corollary \ref{corr:phoneprod-eigenwerte}, there is some nonzero $x\in\C^n$,
such that 
$A_i x = x$ for all $i$. This means, $A_1 \cdots A_N x = x$ and thus
$P_{A_1 \cdots A_N} x = x$.

\qed
\eunum
\epf

\blem
\label{lem:orth_decomp(phone_matrices)}
For any $n$-phone matrices $A_1, \ldots, A_N$ we have:
\bnum2
\item
There are $n'\kreuz n'$ matrices
$A'_1, \ldots, A'_N$, such that for $i = 1, \ldots, N$
\bunum
\item
$A_i = A'_i \dirsum \EINS_l$,
\item
each $A'_i$ is positive hermitian;
\item
$\norm{A'_1 \cdots A'_N} < 1$.
\eunum
Here, $l := \dim \esp_1(A_1 \cdots A_N)$ and $n' := n - l$.
\item
$P_{A_1 \cdots A_N}$ exists and is the 
projector to $\esp_1(A_1 \cdots A_N)$.
\enum
\elem
\bpf
Denote $\esp_1(A_1 \cdots A_N) \teilmenge \C^n$ shortly by $X$. 
By Corollary \ref{corr:phoneprod-eigenwerte}, each
$A_i$ is the identity when restricted to $X$. Since each $A_i$ is
hermitian, $X^\senk$ is preserved by each $A_i$.%
\footnote{Let $x^\senk \in X^\senk$ and $x \in X$. 
Then $\skalprod{x}{A_i x^\senk} = \skalprod{A_i^\ast x}{x^\senk} 
 = \skalprod{A_i x}{x^\senk} = \skalprod{x}{x^\senk} = 0$,
hence $A_i x^\senk \in X^\senk$.}
Hence, we may decompose each $A_i$ into $\EINS_X \dirsum A'_i$ according to
$\C^n = X \dirsum X^\senk$. Here, $A'_i$ is a positive, hermitian 
operator on $X^\senk$. (W.l.o.g., we may assume that $A'_i$ is an
$n' \kreuz n'$ matrix with $n' := n - \dim X$.)
If
$\norm{A'_1 \cdots A'_N}$ was $1$,
then 
\bgl
1 \breitrel= \norm{A'_1 \cdots A'_N} 
  \breitrel\leq \norm{A'_1} \cdots \norm{A'_N} 
  \breitrel\leq 1,
\egl
and each $A'_i$ would be $n$-phone.
Since, however, by construction and by Corollary \ref{corr:phoneprod-eigenwerte}, 
$0 = \esp_1(A'_1) \cap \ldots \cap \esp_1(A'_N) = \esp_1(A'_1 \cdots A'_N)$,
we have $P_{A'_1 \cdots A'_N} = \NULL$, as shown 
in Corollary \ref{corr:phoneprod-projector}.
Consequently, by Corollary \ref{corr:phoneprod-norm1}, 
$\norm{A'_1 \cdots A'_N} \neq 1$.
Obviously, we have $P_{\EINS_X} = \EINS_X$,
such that,
by Lemma \ref{lem:proj_dirsum}, 
$P_{A_1\cdots A_N} = P_{(\EINS_X \dirsum A'_1) \cdots (\EINS_X \dirsum A'_N)}$ 
exists and equals 
$P_{\EINS_X} \dirsum P_{A'_1 \cdots A'_N} = \EINS_X \dirsum \NULL_{X^\senk}$.
It is, of course, hermitian.
\qed
\epf

\bcorr
$\esp_1(A_1 \cdots A_N) = \esp_1(P_{A_1 \cdots A_N})$ 
for any $n$-phone matrices $A_1 \cdots A_N$.
\ecorr


\subsubsection{Norms and Traces}

\blem
\label{lem:norm(ab^ia)_faellt}
Let $A, B$ be $n$-phone matrices. 
Then $\norm{A B^i A} \leq \norm{A B^j A}$ for all $i \geq j$.
\elem
\bpf
Let $D$ be the $n$-phone matrix with $B = D^2$. 
Then 
\bgl
\norm{A B^i A} \breitrel= \norm{(D^i A)^\ast (D^i A)} \breitrel= \norm{D^i A}^2
 \breitrel\leq \norm{D^{i-j}}^2 \norm{D^j A}^2 \breitrel= \norm{A B^j A}.
\egl
\qed
\epf

\bcorr
\label{corr:pbp=0_impliz_pb^ip=0-neu}
Let $A, B$ be $n$-phone matrices. 

Then $A B A = \NULL$ implies $A B^i A = \NULL$ for any $i\in\N_+$.
\ecorr
\bpf
We have $0 = \norm{A B A} \geq \norm{A B^i A} \geq 0$
by Lemma \ref{lem:norm(ab^ia)_faellt}.
\qed
\epf

\blem
\label{lem:proj_sandwich_norm_absch}
For any $n$-phone matrices $A$ and $B$, we have
\bgl
\norm{BAB}^{k+1} \breitrel\leq \norm{(AB^2)^{k}} \breitrel\leq \norm{BAB}^{k-1}.
\egl
If $B$ is even a projector $P$, then
\bgl
\norm{PAP}^{k} \breitrel\leq \norm{(AP)^k} \breitrel\leq \norm{PAP}^{k-1}.
\egl
\elem
\bpf
Since $BAB = B^\ast A B$ is hermitian and positive, we have 
$\norm{(BAB)^k} = \norm{BAB}^k$ for any $k \in \N$. 
Now observe that 
\bgl
\norm{BAB}^{k+1} 
 & = & \norm{(BAB)^{k+1}} 
 \breitrel= \norm{B (AB^2)^k AB} \\
 & \leq & \norm{(AB^2)^k} 
 \breitrel= \norm{AB(BAB)^{k-1}B} \\
 & \leq & \norm{(BAB)^{k-1}}  
 \breitrel= \norm{BAB}^{k-1},
\egl
since $\norm{A} = 1 = \norm{B}$
and, similarly,
\bgl
\norm{PAP}^k 
 & = & \norm{(PAP)^k} 
 \breitrel= \norm{P (AP)^k} 
 \breitrel\leq \norm{(AP)^k} 
 \breitrel= \norm{AP(AP)^{k-1}} \\
 & \leq & \norm{P(AP)^{k-1}}  
 \breitrel= \norm{(PAP)^{k-1}}  
 \breitrel= \norm{PAP}^{k-1} , \\
\egl
since $P^2 = P$ and $\norm{P} = 1$.
\qed
\epf

\bprop
\label{prop:AB=0+aequivalent}
Let $A$ and $B$ be $n$-phone matrices, and let $k\in\N_+$. 
Then
\bgl
AB = \NULL 
 \breitrel\aequ \trm AB = 0
 \breitrel\aequ \trm (AB)^k = 0
 \breitrel\aequ ABA = \NULL.
\egl
\eprop

\bpf
Let $C$ be an $n$-phone matrix with $A = C^2$.
\bunum
\item
First of all, let $\trm (AB)^k = 0$ for some $k \in \N_+$. Since
\bgl
\trm (AB)^k \breitrel= \trm C(CBC)^{k-1} CB \breitrel= \trm (CBC)^k
\egl
and since $CBC$ is positive hermitian,
we have%
\footnote{$\trm D^k = 0$ implies $D = \NULL$ for positive hermitian matrices $D$.}
$CBC = \NULL$ and $ABA = \NULL$.
\item
Next, $ABA = \NULL$ implies $(BA)^\ast BA \ident AB^2 A = \NULL$ by 
Corollary \ref{corr:pbp=0_impliz_pb^ip=0-neu}, whence
$\norm{BA}^2 = \norm{(BA)^\ast BA} = 0$, implying $BA = \NULL$ and $AB = \NULL$.
\item
Finally, of course, $AB = \NULL$ implies $\trm (AB)^k = 0$.
\qed
\eunum
\epf


\subsubsection{Splitting}

\blem
\label{lem:AB=0_iff_decomp}
Let $A$ and $B$ be $n$-phone matrices. Then
$AB = \NULL$ iff there is some $0 < l < n$, some $l$-phone matrix $A'$ and
some $(n-l)$-phone matrix $B'$, such that
\bgl
A = A' \dirsum \NULL_l & \text{ and } & B = \NULL_{n-l} \dirsum B'.
\egl
\elem
Note again, the splitting above means that there is a basis of $\C^n$,
such that $A$ and $B$ can be simultaneously splitted in the way given above.

\bpf
If $A$ and $B$ can be split in the given way, then $AB$ obviously vanishes. 
The other way round, $AB = \NULL$ implies $BA = \NULL$, hence $AB = BA$,
whence $A$ and $B$ can be diagonalized simultaneously.
Now, the statement is trivial.
\qed
\epf

\blem
\label{lem:pbp=0_aequ_tr(pb)=0_aequ_zerlegung}
Let $k \in \N_+$, and let $A$ and $B$ be $n$-phone matrices. 

Then we have 
$\trm (P_A B)^k = 0$
iff there are $0 \leq \alpha < 1$, $0 < l < n$ and
$l$-phone matrices $A'$ and $B'$ with
\bgl
A = \EINS_{n-l} \dirsum \alpha A' & \text{ and } & B = \NULL_{n-l} \dirsum B'.
\egl
\elem
\bpf
By Proposition \ref{prop:AB=0+aequivalent},
$\trm (P_A B)^k = 0$ is equivalent to $P_A B = \NULL$.
Analogously to the proof of Lemma \ref{lem:AB=0_iff_decomp},
we see that, for $P_A B = \NULL$, there is a decomposition 
\bgl
P_A = \EINS_{n-l} \dirsum \NULL_l & \text{ and } & B = \NULL_{n-l} \dirsum B'
\egl
for some $l$-phone matrix $B'$. Since $P_A \neq \EINS$ by $P_A B = \NULL$,
we have $0 < l < n$. Now the implication follows from 
Lemma \ref{lem:decomp_phone_matrix}.
The other direction is trivial.
\qed
\epf


\subsection{Hurwitz Products}
\label{subsect:hurwitzproducts}
In this subsection, 
for completeness, we list several properties of Hurwitz products and their 
traces. The proofs are simple and therefore omitted. They may also be found in \cite{bmv16}.

\bdf
Let $m$ and $k$ be 
integers, and let $A$
and $B$ be $n \kreuz n$ matrices.

Then the \df{Hurwitz product} $S_{m,k}(A,B)$ 
is the sum of all matrix products
containing exactly $m-k$ factors $A$ and $k$ factors $B$.
\edf
For definiteness, we assume the Hurwitz product to be zero 
if $m$, $k$, or $m-k$ is negative.

\blem
\label{lem:real_hurwitz_prod_traces}
The Hurwitz product of any two hermitian $n \kreuz n$ matrices is hermitian.

Consequently, its trace is always real.
\elem

\blem
\label{lem:hurwitzidentity1}
For any $m,k \in \N$ and any hermitian $n \kreuz n$ matrices $A$ and $B$, we have
\bgl
S_{m,k} (A,B) & = & A S_{m-1,k}(A,B) + B S_{m-1,k-1}(A,B), \\
S_{m,k} (A,B) & = & S_{m-1,k}(A,B) A + S_{m-1,k-1}(A,B) B.
\egl
\elem

\blem
\label{lem:hurwitztraceidentity1}
For any $m,k \in \N$ and any hermitian $n \kreuz n$ matrices $A$ and $B$, we have
\bgl
(m-k) \: \trm S_{m,k} (A,B) & = & m \: \trm A S_{m-1,k}(A,B), \\
k \: \trm S_{m,k} (A,B) & = & m \: \trm B S_{m-1,k-1}(A,B).
\egl
\elem

\blem
\label{lem:splitting_smk}
Let $A_i$ and $B_i$ be hermitian $n_i \kreuz n_i$ matrices 
with $n_i \in \N$ for $i = 1,2$. 
Then
\bgl
S_{m,k}(A_1 \dirsum \alpha A_2, B_1 \dirsum \beta B_2)
 & = & S_{m,k}(A_1, B_1) + \alpha^{m-k} \beta^k \: S_{m,k}(A_2, B_2)
\egl
for all $m,k \in \N_+$ with $m > k$ and $\alpha, \beta \in \C$.
\elem


\section{Some Complex Analysis}
\label{sect:complexanalysis}

To prove Theorem \ref{thm:positiv_klein} using complex-analytic
methods we will need to study the behaviour of 
\bgl
\trm \bigl[B (\EINS - \tau A)^{-1}\bigr]^k
\egl\noindent
for $n$-phone matrices $A$ and $B$.
In this section, we provide the necessary statements.
\blem
Let $A$ and $B$ be $n$-phone matrices and let $k \in \N$.
Then we have 
\bgl
(\EINS - \tau A)^{-1} \: \bigl[B (\EINS - \tau A)^{-1}\bigr]^k
 & = & \sum_{\laufm=0}^{\infty} \tau^\laufm \: S_{\laufm+k,k}(A,B)
\egl
for all $\tau \in \C$ with $\betrag \tau < 1$.
\elem
\bpf
Since $\norm A = 1$ and $\betrag \tau < 1$, 
\bgl
(\EINS - \tau A)^{-1} & = & \sum_{\laufm=0}^{\infty} \tau^\laufm  A^\laufm
     \breitrel\ident \sum_{\laufm=0}^{\infty} \tau^\laufm \: S_{\laufm+0,0}(A,B)
\egl
converges absolutely and gives the assertion for $k = 0$.
Inductively, we have 
\newcommand{\platz}{\hspace*{-5pt}}
\bgl[2.7ex]
\platz
 &   & 
       (\EINS - \tau A)^{-1} \: \bigl[B (\EINS - \tau A)^{-1}\bigr]^{k+1} 
 \breitrel=
        \sum_{\laufm=0}^{\infty} \tau^\laufm \: S_{\laufm+k,k}(A,B) \: B \: \sum_{\laufm' = 0}^\infty \tau^{\laufm'} A^{\laufm'} \s
\platz
 & = & \sum_{\laufm,\laufm'=0}^{\infty} \tau^{\laufm + \laufm'} \: \bigl(S_{\laufm+k+1,k+1} (A,B) - S_{\laufm+k,k+1} (A,B) A \bigr) \: A^{\laufm'}  \\
\platz
\platz
 & = & \sum_{\laufm,\laufm'=0}^{\infty} \tau^{\laufm + \laufm'} S_{\laufm+k+1,k+1} (A,B) A^{\laufm'} - \sum_{\laufm=0,\:\laufm'=1}^{\infty}  \tau^{\laufm + \laufm'} S_{\laufm+k+1,k+1} (A,B) A^{\laufm'}   \\
\platz
\platz
 & = & \sum_{\laufm   =0}^{\infty} \tau^{\laufm} S_{\laufm+k+1,k+1} (A,B),
\egl
where we used 
\bgl
S_{\laufm+1,k+1} (A,B) & = & S_{\laufm,k+1} (A,B) A + S_{\laufm,k} (A,B) B
\egl
from Lemma \ref{lem:hurwitzidentity1}
in the second step and 
$S_{k,k+1}(A,B) = 0$
in the third one.
\qed
\epf

\bcorr
\label{corr:hurwitztrace_series}
Let $A$ and $B$ be $n$-phone matrices and let $k \in \N_+$.
Then we have 
\bgl
\trm\bigl[B (\EINS - \tau A)^{-1}\bigr]^k
 & = & \sum_{\laufm=0}^{\infty} \: \tau^\laufm \: \frac{k}{\laufm + k}\: \trm S_{\laufm+k,k}(A,B)
\egl
for all $\tau \in \C$ with $\betrag \tau < 1$.
\ecorr

\bpf
Use the relation $k \: \trm S_{m,k} (A,B) = m \: \trm B S_{m-1,k-1}(A,B)$ 
in Lemma \ref{lem:hurwitztraceidentity1} to derive
\bgl[1ex]
 && \trm \bigl[B (\EINS - \tau A)^{-1}\bigr]^{k} 
 \breitrel= \trm B \: (\EINS - \tau A)^{-1} \: \bigl[B (\EINS - \tau A)^{-1}\bigr]^{k-1} \s
 & = & \sum_{\laufm=0}^{\infty} \tau^\laufm \: \trm B S_{\laufm+k-1,k-1}(A,B) 
 \breitrel= \sum_{\laufm=0}^{\infty} \tau^\laufm \: \frac{k}{\laufm + k} \: \trm S_{\laufm+k,k}(A,B).
\egl
\qed
\epf

\blem
\label{lem:laurent}
Let $k$ be a positive integer, and let
$\rest_\laufj$ be rational holomorphic functions for $\laufj = 0,\ldots,k-1$.
Assume that there is a real $r > 1$, such that 
none of the $\rest_\laufj$ has a pole for $\betrag\tau \leq r$. 
Finally, let $\restkonst_k \in \C$ have positive 
real part and define the analytic function $f$ and its expansion coefficients $f_m$ by
\bgl
f(\tau) & \ident & 
\sum_{m = 0}^\infty f_m \tau^m
 \breitrel{:=} \frac{\restkonst_k}{(1 - \tau)^k}  
          + \sum_{\laufj=0}^{k-1} \frac{\rest_\laufj(\tau)}{(1 - \tau)^\laufj}
\egl

Then there is an $m_0 \in \N$, such that $\re f_m > 0$ for all $m \geq m_0$.
\elem
\bpf
Write $f$ as a Laurent series 
\bgl
f(\tau) 
 & = & \frac{\restkonst_k}{(1 - \tau)^k}  + \sum_{\laufj=0}^{k-1} \frac{\restkonst_\laufj}{(1 - \tau)^\laufj} + \Rest(\tau)
\egl
around $1$ for appropriate $\restkonst_\laufj \in \C$ and 
holomorphic $\Rest$. Then $\Rest$ is again rational and has no pole for 
$\betrag \tau \leq r$. 
Consequently, the norm of the $\laufm$-th Taylor coefficient of
$\Rest(\tau)$ can be estimated by 
$\frac{\Restkonst}{r^\laufm}$
for some constant $\Restkonst \geq 0$. 
As, by Lemma \ref{lem:eulerseries}, the $m$-th Taylor coefficient of
\bgl
\inv{(1-\tau)^{\laufj}} & = & \sum_{\laufm = 0}^\infty \binom{\laufm + \laufj - 1}{\laufm} \: \tau^\laufm
\egl
is a polynomial in $\laufm$ of degree $\laufj - 1$ with leading coefficient $\inv{(\laufj-1)!}$, 
the assertion is obvious since $\re\restkonst_k > 0$.
\qed
\epf


\section{Some Combinatorics}
\label{sect:combinatorics}
The ultimate goal of this article is to derive 
asymptotic properties of $\trm S_{m,k}(A,B)$.
Recall that $S_{m,k}(A,B)$ equals the sum of 
all products of matrices where $m-k$ factors
equal $A$ and $k$ factors equal $B$.
The trace of such a single product significantly depends on its
``factor pattern''. For instance, if the substring $AB$ appears $l$ times
in the matrix product, then the trace of the full product 
cannot exceed $n \norm{AB}^l$.
To finally estimate the sum of all these product traces, we
need estimates how frequently this pattern appears.
This now is a purely combinatorial problem for words in two letters.
To avoid confusion we will denote the letters by $a$ and $b$, and return to $A$ and $B$ only later.
Let, moreover, $0\leq k\leq m$ be integers and
denote the 
set of all words containing exactly $m-k$ letters $a$ and $k$ letters
$b$ by $\word_{m,k}$.


\subsection{Counting}

\bprop
\label{prop:anz_s_subwords_ab}
Denote by $\wechsel_{m,k,s} \teilmenge \word_{m,k}$ the set of words
containing exactly $s$ times the subword $ab$.
Then we have 
\bgl
\betrag{\wechsel_{m,k,s}} & = & \binom{m-k}s \binom ks
\egl
and
\bgl
\betrag{\word_{m,k}} & = & \sum_{s} \betrag{\wechsel_{m,k,s}} \breitrel= \binom{m}k .
\egl

\eprop
Here, we used the convention that $\binom ij = 0$ if $j>i$.
\bpf
Let $w\in\wechsel_{m,k,s}$ be a word with exactly $s$ subwords $ab$. 
Then 
\bgl
w & = & b^{j_0} a^{i_1} b^{j_1} \cdots a^{i_s} b^{j_s} a^{i_{s+1}}
\egl
for appropriate $i_\iota,j_\iota \geq 1$, $\iota=1,\ldots,s$, 
and $j_0, i_{s+1} \geq 0$
with $i_1 + \ldots + i_{s+1} = m-k$ and $j_0 + \ldots + j_s = k$. 
Obviously, it is sufficient to prove 
that there are exactly $\binom ks$ ways to write $k$ as a sum 
$j_0 + j_1 + \ldots + j_s$ of 
$s+1$ integers with $j_0 \geq 0$ and $j_\iota \geq 1$.
In fact, there are $\binom ks$ possibilities to choose 
$s$ elements $J_1 < \ldots < J_{s}$ out of the $k$ numbers $1,\ldots,k$.
Letting $j_0 := J_1 - 1$ and $j_\iota := J_{\iota+1} - J_{\iota}$
for $0 < \iota < s$
and $j_s := k+1 - J_s$ gives such a decomposition $j_0 + j_1 + \ldots + j_s$
of $k$. Since, the other way round, each such decomposition can be 
obtained by such $J_\iota$, we get the proof.

The second assertion is clear.
\qed
\epf

\blem
\label{lem:woerterzahl_mit_k_wechseln}
\bnum2
\item
No word in $\wechsel_{m,k,k}$ contains the subword $bb$,
i.e., any word in $\wechsel_{m,k,k}$ can be written as
$a^{i_1} b a^{i_2} \cdots a^{i_k} b a^{i_{k+1}}$ 
with $i_\iota > 0$ and $i_{k+1} \geq 0$.
\item
Denote by $\kwechsel_{m,k,\minimallaenge} \teilmenge \wechsel_{m,k,k}$ the set of 
those words $a^{i_1} b a^{i_2} \cdots a^{i_k} b a^{i_{k+1}}$ as above
with $i_\iota > \minimallaenge$ and $i_{k+1} \geq \minimallaenge$
for some integer $\minimallaenge \geq 0$. 
Then 

\bgl
\betrag{\kwechsel_{m,k,\minimallaenge}} & = & 
\betrag{\wechsel_{m-(k+1)\minimallaenge,k,k}}.
\egl
\enum
\elem
\bpf
\bnum2
\item
This follows directly from the proof of 
Proposition \ref{prop:anz_s_subwords_ab}.
In fact, let
\bgl
w & = & b^{j_0} a^{i_1} b^{j_1} \cdots a^{i_k} b^{j_k} a^{i_{k+1}} 
   \breitrel\in \wechsel_{m,k,k}.
\egl 
Since $j_0 + \ldots + j_k = k$ and $j_1, \ldots, j_k > 0$, we have
$j_0 = 0$ and $j_1 = \ldots = j_k = 1$. 
\item
One easily checks that
\fktdefabgesetzt{\irgendeinefunktion}%
   {\wechsel_{m-(k+1)\minimallaenge,k,k}}%
   {\kwechsel_{m,k,\minimallaenge} \teilmenge \wechsel_{m,k,k}.}%
   {a^{i_1} b a^{i_2} \cdots a^{i_k} b a^{i_{k+1}}}
   {a^{i_1+\minimallaenge} b a^{i_2+\minimallaenge} \cdots a^{i_k+\minimallaenge} b a^{i_{k+1}+\minimallaenge}}
is a bijection.
\qed
\enum
\epf


\subsection{Estimates}

We will need estimates on how the number of words changes 
in the event of having a fixed amount of letters less and 
how often there are subwords $ab$.
In the first simple lemma, we will see that 
the (relative) decrease of the word number while
dropping a finite number of letters $a$ is arbitrarily small provided 
we had started with $a$ occurring sufficiently often.
In the second lemma, we show that ---again for $a$ occurring sufficiently
often, i.e., for large $m$--- the (relative) 
number of words containing less than $k$ subwords
$ab$ can be made arbitrarily small. Or, in other words, 
if one of the $k$ letters $b$ appears then it appears ``lonely'', i.e.,
$b^2$ or higher powers typically do not appear.

\neueseite

\blem
\label{lem:binom_zaehleraenderung_absch}
For $0 < \varepsilon < 1$, positive integers $\minimallaenge$ and $m$ 
with $m \geq \minimallaenge(1 + \frac{k}{\varepsilon})$, we have
\bgl
\binom{m-\minimallaenge}k & \geq & (1-\varepsilon) \: \binom{m}k.
\egl
\elem
\bpf
Use
\bgl
\hspace*{-2ex}
\binom{m-\minimallaenge}k 
 \breitrel= \binom{m}k  \prod_{j=0}^{\minimallaenge-1} \Bigl(1 - \frac k{m-j} \Bigr) 
 \breitrel\geq \binom{m}k \prod_{j=0}^{\minimallaenge-1} \Bigl(1 - \frac{\varepsilon}\minimallaenge\Bigr) 
 \breitrel\geq \binom{m}k (1 - \varepsilon).
\hspace*{-2ex}
\egl
\qed
\epf

\blem
\label{lem:absch_woerterzahl_mit_weniger_als_S_abs}
Let $0 < \varepsilon < 1 \leq S$ and 
\bgl
m \breitrel> \frac{S^3}\varepsilon + 2S -1 
& \text{ and } & 
k, m-k  \breitrel\geq S. \\
\egl
Then 
\bgl
\sum_{s = 0}^{S-1} \binom{m-k}s \binom ks 
  & < & \varepsilon\binom{m-k}S \binom kS.
\egl
\elem
\bpf
Observe that for $0 \leq s \leq S \leq k,m-k$ and for $m$ as in the lemma
\bgl[1.5ex]
\frac{s^2}\varepsilon 
  & \leq & \frac{S^3}\varepsilon 
  \breitrel\leq \frac{S^3}\varepsilon + 2S -1 - (2s-1) 
  \breitrel< m - 2s +1 
   \s
  & \leq & ((m-k) -s)(k-s) + m - 2s +1 
  \breitrel=  ((m-k) -s + 1)(k-s + 1) .
\egl
Using the abbreviation $\doppelbinom_s := \binom{m-k}s \binom ks$ for all
$s \in \N$, one immediately checks that
\bgl
\doppelbinom_{s-1}
 & = & \frac{s^2}{(m-k-s+1)(k-s+1)} \:\: \doppelbinom_{s}.
\egl
As just seen above, the prefactor is always smaller than $\varepsilon < 1$, whence we get
\bgl
\sum_{s=0}^{S-1} d_s 
  & \leq & \sum_{s=0}^{S-1} d_{S-1} 
  \breitrel= S d_{S-1} 
  \breitrel= \frac{S^3}{(m-k-S+1)(k-S+1)} \:\: d_{S} 
  \breitrel< \varepsilon \: d_{S}. \\
\egl
\qed
\epf



\section{Proofs of the Main Theorems}

\label{sect:main_thms}

\subsection{Reduction of Theorem \ref{thm:positiv_klein} to the Case $P_A B \neq \NULL$}

\bprop
\label{prop:reduct_PABneq0}
If the assertions of Theorem \ref{thm:positiv_klein} hold for 
$m > k > 0$ and for any
$n$-phone matrices $A$ and $B$ with $P_A B \neq \NULL$, 
then Theorem \ref{thm:positiv_klein} is valid in toto.
\eprop
\bpf
For the proof of Theorem \ref{thm:positiv_klein}, 
we may, of course, assume that $A$ and $B$ are $n$-phone matrices.
Moreover, we may assume that $m > k > 0$, that $n > 1$ and that $AB \neq \NULL$,
as the other cases are trivial.

Assume now that $P_A B = \NULL$. 
Then, by Lemma \ref{lem:pbp=0_aequ_tr(pb)=0_aequ_zerlegung},
we find some $0 \leq \alpha < 1$ and some $l$-phone matrices $A'$ and $B'$
with $0 < l < n$, such that 
\bgl
A = \EINS_{n-l} \dirsum \alpha A' 
        & \text{ and } & B = \NULL_{n-l} \dirsum B'.
\egl
Since 
$\NULL \neq AB = \NULL_{n-l} \dirsum \alpha A' B'$,
we have $\alpha \neq 0$ and $A'B' \neq \NULL_l$.
Together with 
\bgl
\hspace*{-2ex}
S_{m,k}(A,B) 
 & = & S_{m,k}(\EINS_{n-l},\NULL_{n-l}) 
                      + \alpha^{m-k} \:  S_{m,k}(A',B') 
 \breitrel= \alpha^{m-k} \:  S_{m,k}(A',B')
\hspace*{-2ex}
\egl
by Lemma \ref{lem:splitting_smk} and $m > k > 0$, this implies the
assertion by induction. 
\qed
\epf

\subsection{Complex-Analytic Proof of Theorem \ref{thm:positiv_klein}}

First we prove our Main Theorem 
 by means of complex analysis
without focussing on 
concrete estimates.

\bpf[Theorem \ref{thm:positiv_klein}]
Let $A$ and $B$ be $n$-phone matrices.
Using Lemmata \ref{lem:ew1} and \ref{lem:potenz_diff_n-phone_proj}, 
we have
\bgl
P_A (\EINS - \tau A)^{-1} 
  & = & \sum_{\laufm=0}^\infty P_A \: \tau^\laufm A^\laufm
  \breitrel= \sum_{\laufm=0}^\infty \tau^\laufm \: P_A 
  \breitrel= \inv{1 - \tau} \: P_A 
\egl
and
\bgl
(\EINS - P_A) (\EINS - \tau A)^{-1} 
  & = & \sum_{\laufm=0}^\infty (\EINS - P_A) \: \tau^\laufm A^\laufm \\
  & = & \sum_{\laufm=0}^\infty \tau^\laufm \: (A^\laufm - P_A)
  \breitrel= \sum_{\laufm=0}^\infty \tau^\laufm \: (A - P_A)^\laufm.
\egl
Observe that 
$(\EINS - P_A) (\EINS - \tau A)^{-1}$ is rational 
and (up to the removable discontinuity at $1$) analytic for $\betrag \tau < \norm{A - P_A}^{-1}$,
whereas $\norm{A - P_A}^{-1}$ is strictly larger than $1$.
From
\bgl[1ex]
B (\EINS - \tau A)^{-1} 
  & = & B \bigl(P_A + (\EINS - P_A)\bigr) (\EINS - \tau A)^{-1} \s
  & = & \inv{1 - \tau} \: B P_A + B (\EINS - P_A) (\EINS - \tau A)^{-1} ,
\egl
we derive
\bgl
\trm\bigl[B (\EINS - \tau A)^{-1}\bigr]^k
 & = & \frac{\trm (P_A B)^k}{(1 - \tau)^k}  
          + \sum_{\laufj=0}^{k-1} \frac{\rest_\laufj(\tau)}{(1 - \tau)^\laufj} \, ,
\egl
whereas each $\rest_\laufj$ is a rational function vanishing at infinity and
having poles only for
\bgl
\betrag \tau 
  & \geq & \inv{\norm{A - P_A}}
  \breitrel> 1.
\egl  
Now the assertion follows from Lemma \ref{lem:laurent} together with
Corollary \ref{corr:hurwitztrace_series}: In fact,
we may assume $P_A B \neq \NULL$, hence $\trm P_A B > 0$, 
as well as $k > 0$ by Proposition \ref{prop:reduct_PABneq0},
and we know that any Hurwitz product trace is real by Lemma \ref{lem:real_hurwitz_prod_traces}.
\qed
\epf


\subsection{Combinatorial Proof of Theorem \ref{thm:positiv_klein}}
There are two main steps
in the study of the asymptotics 
of $\trm S_{m,k}(A,B)$ 
for growing $m$ while $k$ is fixed:
First, we estimate how fast the
products $A^{l_1} B \cdots A^{l_k} B$ do approach $(P_A B)^k$
depending on a lower bound $\minimallaenge$ to all $l_i$.
Second, the longer the words are (i.e., for growing $m$), 
all other words (i.e., those with $l_i < \minimallaenge$ or having substrings $b^2$)
get less frequent for fixed $\minimallaenge$.
This allows us 
to estimate how fast $\trm S_{m,k}(A,B) / \trm S_{m,k}(\EINS,\EINS)$ approaches
$\trm(P_A B)^k / n$ and, finally, to prove Theorem \ref{thm:positiv_klein}.

\blem
\label{lem:absch_prod_AkB_minus_prod_P_A_B}
For any $n$-phone matrices $A$ and $B$ and for any
integers $l_1, \ldots, l_k \geq \minimallaenge > 0$, we have

\bgl
\Bignorm{\prod_{i=1}^k A^{l_i} B - \prod_{i=1}^k P_A B}
 & \leq & k \: \norm{A - P_A}^{\minimallaenge} \: \norm{A^\minimallaenge B}^{k-1}.
\egl
\elem
\bpf
Observe that for any $n \kreuz n$ matrices $X_1, \ldots, X_k$ and $X$, we have
(see Lemma \ref{lem:prod_diff_zerlegung})
\bgl
X_1 \cdots X_k
 & = & X^k + \sum_{i=1}^{k} X^{i-1} \: (X_i - X) \: X_{i+1} \cdots X_k.
\egl
Now, Lemma \ref{lem:potenz_diff_n-phone_proj} 
implies
\bgl[3ex]
\hspace*{-1ex}
\Bignorm{\prod_{i=1}^k A^{l_i} B - \prod_{i=1}^k P_A B }
 & \leq &  \sum_{i=1}^{k} 
           \norm{(P_A B)^{i-1}} \: \norm{(A^{l_i} - P_A) B} \: \norm{A^{l_{i+1}} B} \cdots \norm{A^{l_k} B} \\
\hspace*{-1ex}
 & \leq &  \sum_{i=1}^{k} 
           \norm{P_A B}^{i-1} \: \norm{A - P_A}^{l_i} \: \norm{A^{l_{i+1}} B} \cdots \norm{A^{l_k} B} \\
 & \leq &  \norm{A - P_A}^{\minimallaenge} \sum_{i=1}^{k} 
           \norm{A^\minimallaenge B}^{i-1} \: \norm{A^\minimallaenge B}^{k-i} \s
 & = &  k \: \norm{A - P_A}^{\minimallaenge} \: \norm{A^\minimallaenge B}^{k-1}.
\egl
\qed
\epf
In Section \ref{sect:combinatorics}, we studied words in the two letters $a$ 
and $b$. We now define $\wmers$ to be the homomorphism from 
$\word_{m,k}$ to the $n \kreuz n$ matrices, whereas
$\wmers(a) := A$ and $\wmers(b) := B$. It is now clear that, e.g.,
$S_{m,k} (A,B) = \sum_{w\in\word_{m,k}} \wmers(w)$.

\bprop
\label{prop:k_klein_absch}
Let $A$ and $B$ be $n$-phone matrices, and 
let $k \in \N$ and $\varepsilon \in (0,1)$ be fixed.
Choose now some $\minimallaenge \in \N_+$, such that 
\bgl
k \: \norm{A - P_A}^{\minimallaenge} \: \norm{A^\minimallaenge B}^{k-1} & < &  \varepsilon.
\egl
Then, for any $m \in \N$ with
\bgl
m & \geq & \Bigl(1 + \frac{k}\varepsilon\Bigr) \: \bigl(k+k\minimallaenge+\minimallaenge\bigr),
\egl
we have
\bgl
\Bigbetrag{\frac{\trm S_{m,k}(A,B)}{\trm S_{m,k}(\EINS,\EINS)} - \frac{\trm (P_A B)^k}n}
 & \leq & \Bigl(\frac{\trm (P_A B)^k}n + 2\Bigr) \: \varepsilon.
\egl
\eprop
Observe that 
$\trm (P_A B)^k$ 
is always nonnegative. 
\bpf
First observe, that $\norm{A - P_A} < 1$ 
by Corollary \ref{corr:norm(a-pa)_kleiner_1},
whence there exists such an $\minimallaenge$. Next, observe that
\bgl[2.5ex]
\betrag{\word_{m,k} \setminus \kwechsel_{m,k,\minimallaenge}} 
 & = & \betrag{\word_{m,k}} - \betrag{\kwechsel_{m,k,\minimallaenge}} \s
 & = & \betrag{\word_{m,k}} - \betrag{\wechsel_{m-(k+1)\minimallaenge,k,k}} \erl{by Lemma \ref{lem:woerterzahl_mit_k_wechseln}}\s
 & = & \binom{m}{k} - \binom{m-(k+1)\minimallaenge-k}{k} \binom{k}{k} \erl{by Proposition \ref{prop:anz_s_subwords_ab}}\s
 & \leq & \binom{m}{k} - (1-\varepsilon) \: \binom{m}{k} \erl{by Lemma \ref{lem:binom_zaehleraenderung_absch}}\s
 & = & \varepsilon\betrag{\word_{m,k}}. 
\egl
since $m \geq \bigl((k+1)\minimallaenge+k\bigr) \bigl(1+\frac{k}\varepsilon\bigr)$ by assumption.
Third,
observe that for $w \in \kwechsel_{m,k,\minimallaenge}$, we have

\bgl
\betrag{\trm \wmers(w) - \trm (P_A B)^k} & \leq & n k \: \norm{A - P_A}^{\minimallaenge} \: \norm{A^\minimallaenge B}^{k-1} \breitrel< n\varepsilon
\egl
by Lemma \ref{lem:absch_prod_AkB_minus_prod_P_A_B} and $\betrag{\trm C} \leq n \norm{C}$ for any
matrix $C$, whence

\bgl
 && \Bigbetrag{%
    \frac{\sum_{w\in\word_{m,k}} \trm \wmers(w)}{\trm S_{m,k}(\EINS,\EINS)} - \frac{\trm (P_A B)^k}n }
\\
 & = & \Bigbetrag{\sum_{w\in\word_{m,k}} 
    \frac{\trm \wmers(w)-\trm (P_A B)^k}{\trm S_{m,k}(\EINS,\EINS)} } \\
 & \leq & \Bigbetrag{\sum_{w\in\kwechsel_{m,k,\minimallaenge}} 
               \frac{\trm \wmers(w) - \trm (P_A B)^k}{\trm S_{m,k}(\EINS,\EINS)}}
          \: + \: \sum_{w\in\word_{m,k} \setminus \kwechsel_{m,k,\minimallaenge}} \Bigbetrag{
               \frac{\trm \wmers(w) - \trm (P_A B)^k}{\trm S_{m,k}(\EINS,\EINS)}} \\
 & < & \frac{\betrag{\kwechsel_{m,k,\minimallaenge}}}{n\betrag{\word_{m,k}}} \: n\varepsilon
          \: + \: \frac{\betrag{\word_{m,k}}-\betrag{\kwechsel_{m,k,\minimallaenge}}}{n\betrag{\word_{m,k}}}
	  \: \bigl(n + \betrag{\trm (P_A B)^k}\bigr) 
	  \\
 & < & 
         \Bigl(2 + \frac{{\trm (P_A B)^k}}n\Bigr) \: \varepsilon 
\egl
using
\bgl
\betrag{\word_{m,k}} 
 \breitrel= \binom mk 
 \breitrel= \sum_{w\in\word_{m,k}} 1 
 \breitrel= \inv n \: \trm S_{m,k}(\EINS,\EINS).
\egl
\qed
\epf
Theorem \ref{thm:positiv_klein} is now a corollary:

\bpf[Theorem \ref{thm:positiv_klein}]
By Proposition \ref{prop:reduct_PABneq0}, we may assume 
that $k > 0$ and that $A$ and $B$ are $n$-phone matrices with
$P_A B \neq \NULL$ or,
equivalently, with $\trm(P_A B)^k \neq 0$ by Proposition \ref{prop:AB=0+aequivalent}.
Then there are $\minimallaenge > 0$ and $\varepsilon\in(0,1)$ with
\bgl
k \: \norm{A - P_A}^{\minimallaenge} \: \norm{A^\minimallaenge B}^{k-1} 
  \breitrel<  \varepsilon
  \breitrel<  \frac{\trm (P_A B)^k}{3n}\:.
\egl
Now, since $0 < \trm (P_A B)^k \leq n$, we have
\bgl[1.8ex]
\frac{\trm S_{m,k}(A,B)}{\trm S_{m,k}(\EINS,\EINS)} 
 & \geq & \frac{\trm (P_A B)^k}n - \Bigl(\frac{\trm (P_A B)^k}n + 2\Bigr) \: \varepsilon \\
 & > & \frac{\trm (P_A B)^k}n - \Bigl(\frac{\trm (P_A B)^k}n + 2\Bigr) \: \frac{\trm (P_A B)^k}{3n} \s
 & = & \inv3 \frac{\trm (P_A B)^k}n \Bigl(1-\frac{\trm (P_A B)^k}n\Bigr) 
 \breitrel\geq 0,
\egl
provided
\bgl
m & \geq & m_0 \breitrel{:=} \Bigl(1 + \frac{k}\varepsilon\Bigr) \: \bigl(k+k\minimallaenge+\minimallaenge\bigr).
\egl
\qed
\epf

\brem
The proof of Theorem \ref{thm:positiv_klein} above provides us with an
explicit estimate for the value of $m_0$. 
If $A$ is not a projector and $P_A B \neq \NULL$, 
we have $\trm S_{m,k}(A,B) > 0$
for all $m \geq m_0$ with
\bgl
m_0 
 & := & (1 + k) \: \Bigl(1 + \frac{3kn}{\trm (P_A B)^k}\Bigr) \: 
         \Bigl(2+ \frac{\ln \trm (P_A B)^k - \ln 3kn}{\ln \norm{A - P_A}} \Bigr).
\egl
If $A$ is a projector 
and $A B \neq \NULL$, 
then $\trm S_{m,k}(A,B) > 0$
for all $m \geq m_0$ with
\bgl
m_0 
 & := & (1 + 2k) \: \Bigl(1 + \frac{3kn}{\trm (A B)^k}\Bigr).
\egl
For $P_A B = \NULL$, use the decompositions of $A$ and $B$ 
as in the proof of Proposition \ref{prop:reduct_PABneq0} 
and then use the expressions above with $A$ and $B$ replaced
by $A'$ and $B'$, respectively. (If again $P_{A'} B' = \NULL$,
proceed iteratively.)
Of course, the estimates above need not be optimal; if the BMV conjecture
was true, $m_0$ would probably be $k$ unless $AB = 0$.
\erem


\subsection{Growing $m$ and Not-too-small $k$}
If $ab$ appears $S$ times in a word in $\word_{m,k}$,
then the corresponding matrix product has at most norm $\norm{AB}^S$.
For growing $m$, the typical number of alternations
between $a$ and $b$ in a word indeed increases; in particular, 
it passes the threshold $S$ sooner or later. 
Therefore, the normalized trace of $S_{m,k}(A,B)$ 
can be estimated by $\norm{AB}^S$ up to some $\varepsilon$.
\bprop
\label{prop:k_nichtklein_absch}
Let $0 < \varepsilon < 1 \leq S$ for some $S \in \N$ and 
\bgl
m \breitrel> \frac{S^3}\varepsilon + 2S -1 
& \text{ and } & 
k, m-k  \breitrel\geq S.
\egl
Then 
\bgl
\Bigbetrag{\frac{\trm S_{m,k}(A,B)}{\trm S_{m,k}(\EINS,\EINS)}}
 & < & \varepsilon + \norm{AB}^S.
\egl
\eprop
\bpf
First observe that 
$\betrag{\trm \wmers(w)} \leq n \norm{\wmers(w)} \leq n \norm{AB}^S$ 
for any $w \in \kwechsel_{m,k,s}$ with $s \geq S$. 
Now, we simply decompose all elements of $\word_{m,k}$
into two sets: one consisting of all elements containing
less then $S$ subwords $ab$ and the other one consisting of
all elements with at least $S$ subwords $ab$. We get
\bgl[1.8ex]
\betrag{\trm S_{m,k}(A,B)}
 & \leq & \sum_{s < S} \sum_{w \in \wechsel_{m,k,s}} \betrag{\trm\wmers(w)}
          + \sum_{s \geq S} \sum_{w \in \wechsel_{m,k,s}} \betrag{\trm\wmers(w)} \\
 & \leq & \sum_{s < S} \betrag{\wechsel_{m,k,s}} \: n
          + \sum_{s \geq S} \sum_{w \in \wechsel_{m,k,s}} n \: \norm{AB}^S  \\
 & < & \varepsilon \: \betrag{\wechsel_{m,k,S}} \: n
          + \betrag{\word_{m,k}} \: n \: \norm{AB}^S
	  \erl{by Lemma \ref{lem:absch_woerterzahl_mit_weniger_als_S_abs}}\s
 & \leq & n \: \betrag{\word_{m,k}}  (\varepsilon + \: \norm{AB}^S) \s
 & \leq & \trm S_{m,k}(\EINS,\EINS) \: (\varepsilon + \: \norm{AB}^S). 
\egl
\qed
\epf


\subsection{Asymptotics for Growing $m$ and General $k$}
\newcommand{\teil}{2}

Since $\esp_1(A) \cap \esp_1(B) = \esp_1(AB)$,
Theorem \ref{thm:asymptotik} follows immediately from 
\bthm
For any $n$-phone matrices $A$ and $B$, and for any $0 < \varepsilon < 1$, 
there are some $m_0 \in \N$
and some $k_0 \in \N$, 
such that for all $m \geq m_0$ 
\bgl
\frac{\trm S_{m,k}(A,B)}{\trm S_{m,k}(\EINS,\EINS)}
 & > & \frac{\dim \esp_1(AB)}{n} - \varepsilon 
              \text{ \:\:\: for all $0 \leq k \leq m$ }
\egl
and 

\bgl
\frac{\trm S_{m,k}(A,B)}{\trm S_{m,k}(\EINS,\EINS)}
 & < & \frac{\dim \esp_1(AB)}{n} + \varepsilon 
              \text{ \:\:\: for all $k_0 \leq k \leq m - k_0$ }.
\egl

\ethm
\newcommand{\kzero}{{k_0}}
\bpf
First of all, let us find $\kzero$ and $m'_0$, such that
\bgl
\Bigbetrag{\frac{\trm S_{m,k}(A,B)}{\trm S_{m,k}(\EINS,\EINS)} - \frac{\dim \esp_1(AB)}{n}}
 & < &  \varepsilon 
\egl
for all $m \geq m'_0$ and $k_0 \leq k \leq m - k_0$.
\bunum
\item
Assume first $\norm{AB} < 1$, i.e.,
$\esp_1(AB) = 0$ by Corollary \ref{corr:phoneprod-norm1}.
Choose some positive integer $\kzero$,
such that
\bgl
\norm{AB}^\kzero & < & \frac{\varepsilon}\teil \: ,
\egl
\neueseite
and some $m'_0 \in \N$, such that 
\bgl
m'_0 & > & \frac{\teil \kzero^3}\varepsilon + 2\kzero - 1.
\egl
Now, Proposition \ref{prop:k_nichtklein_absch} implies
that 
\bgl
\Bigbetrag{\frac{\trm S_{m,k}(A,B)}{\trm S_{m,k}(\EINS,\EINS)}}
 & < & \varepsilon
\egl
for all $m \geq m'_0$ and all $\kzero \leq k \leq m - \kzero$.
\item
Assume now $\norm{AB} = 1$. According to 
Lemma \ref{lem:orth_decomp(phone_matrices)},
we may decompose $A$ and $B$ into $A = A' \dirsum \EINS_l$ and
$B = B' \dirsum \EINS_l$ with $\norm{A' B'} < 1$ for 
$l := \dim \esp_1 (AB)$. Using Lemma \ref{lem:splitting_smk}, 
we have
\bgl
\trm S_{m,k} (A,B) 
   & = & \trm S_{m,k}(A',B') 
           + \trm S_{m,k} (\EINS_l,\EINS_l),
\egl
and, therefore,
\bgl
\frac{\trm S_{m,k} (A,B)}{\trm S_{m,k} (\EINS_n,\EINS_n)} - \frac ln  
   & = & \frac ln \: 
          \frac{\trm S_{m,k} (A',B')}{\trm S_{m,k} (\EINS_l,\EINS_l)}.
\egl
\bunum
\item
If $\norm{A'} \norm{B'} = 1$, then
$A'$ and $B'$ are $l$-phone matrices with $\norm{A' B'} < 1$,
for that the result has been established above.
\item
If $\norm{A'} \norm{B'} < 1$, then 
choose $\kzero \in \N$, 
such that $(\norm{A'} \norm{B'})^\kzero < \varepsilon$.
Then,
by Lemma \ref{lem:splitting_smk},
we have 
\bgl
\Bigbetrag{\frac{\trm S_{m,k} (A,B)}{\trm S_{m,k} (\EINS_n,\EINS_n)} - \frac ln}
   & \leq & \frac l n \: \norm{A'}^{m-k} \norm{B'}^k 
   \breitrel< \varepsilon
\egl
for any $m,k \in \N$ with $m-k \geq \kzero$ and $k \geq \kzero$.
\eunum

\eunum

Now, let us finish the proof by showing that
\bgl
\frac{\trm S_{m,k}(A,B)}{\trm S_{m,k}(\EINS,\EINS)} - \frac{\dim \esp_1(AB)}{n}
 & \geq &  0
\egl
for all $m \geq m_0$ with an appropriate $m_0$ 
and for $k \leq k_0$ or $k \geq m - k_0$.
\bunum
\item
Assume again first that $\norm{AB} < 1$.
Now, according to Theorem \ref{thm:positiv_klein}, 
for each $k \in \N$, there is some integer
$m'_0(k)$, such that 
\bgl
\trm S_{m,k}(A,B) \breitrel\geq 0
 & \text{ and } & 
\trm S_{m,m-k}(A,B) \breitrel\ident \trm S_{m,k}(B,A) \breitrel\geq 0
\egl
for all $m \geq m'_0(k)$.
Now, simply define
\bgl
m_0 & := & \max_{k \leq \kzero} \big\{m'_0(k), m'_0\big\},
\egl
and we have the desired assertion.
\item
If $\norm{AB} = 1$, we decompose $A$ and $B$ as above into
$A = A' \dirsum \EINS_l$ and $B = B' \dirsum \EINS_l$ 
with $\norm{A' B'} < 1$. 
Again using Lemma \ref{lem:splitting_smk},
the assertion follows as in the previous case.
\qed
\eunum

\epf


\section*{Acknowledgements}
The authors are grateful to Julius Borcea, who informed Shmuel Friedland about 
the paper \cite{paper27} of Christian Fleischhack and this way ultimately 
triggered the present joint paper.
Christian Fleisch\-hack is very grateful for the kind hospitality 
granted to him by the Max-Planck-Insti\-tut f\"ur Mathe\-matik in den Natur\-wissen\-schaf\-ten
in Leipzig as well as by the Institut f\"ur Theoretische Physik at the Uni\-ver\-si\-t\"at Leip\-zig.
Moreover, he thanks Johannes Brunne\-mann and Ulf K\"uhn 
for discussions.
The work has been supported in part by the Emmy-Noether-Programm 
(grant FL~622/1-1) of the Deut\-sche For\-schungs\-gemein\-schaft.


\anhangengl

\section{Euler-Lagrange Equations}
\label{app:eleq}
In the main body of the article, we have always used the operator
norm for matrices and reduced our investigations typically to 
normalized matrices. In fact, this has been justified by the homogeneity
of $S_{m,k}(\cdot,\cdot)$. Nevertheless, there is a 
full range of other possible norms that can be taken to
normalize the matrices. In \cite{bmv16}, e.g., the Frobenius
norm has been used to derive the Euler-Lagrange equations
of the BMV conjecture. They yielded, among others, relations between 
$A^2$ and $A S_{m-1,k}(A,B)$ in any point where $\trm S_{m,k}$
is minimal or maximal.
In this appendix we are going to rederive these relations
in a slightly more abstract way and extend them to other norms.

\newcommand{\nl}{p}
\newcommand{\su}{\mathfrak{su}}
For that purpose, we choose the following Schatten $\nl$-norms%
\footnote{Note, that the Schatten $\nl$-norm is actually defined \cite{Bhatia} by 
\bgl
\Bigl[\sum_{i=1}^n  s_j(A)^\nl\Bigr]^{\inv\nl}, 
\egl\noindent
where $s_j(A)$, $j = 1, \ldots, n$, are the singular values
of $A$. For positive hermitian
matrices, however, our notion coincides with that definition.
As we are interested in the case of positivity only, we
may sloppily re-use the notion $\nl$-``norm'' for our case.
In fact, our definition does not give a norm on the linear space of all $n \kreuz n$ matrices.}
\bgl
\norm{A}_\nl := \sqrt[\nl]{\trm A^\nl}
& \text{ and } &
\norm{A}_\infty := \norm{A}
\egl\noindent
for $\nl \geq 1$ and for positive hermitian $A$.
One immediately sees that 
$\norm{A}_\infty = \lim_{\nl\gegen\infty} \norm{A}_\nl$.
Let us now fix some $\nl \in [1,\infty]$. Moreover, to avoid cumbersome
notation, we let $n$-phone matrices be positive hermitian matrices
having $\nl$-norm $1$ (instead of to be of operator norm $1$
as in the main text).
Next, observe that for any matrix-valued functions $f$ and $g$ on $\R$,
we have
\bgl
\frac{\dd}{\dd x} \: \trm (f + tg)^m 
 & = & m \: \trm \Bigl(\frac{\dd(f+tg)}{\dd x} \: (f + tg)^{m-1} \:\Bigr)
\egl\noindent
and, by comparison of coefficients,
\bgl
\trm S'_{m,k} (f,g) 
 & = & m \: \trm 
        \bigl(f' \: S_{m-1,k}(f,g) + g' \: S_{m-1, k-1}(f,g) \bigr).
\egl\noindent
Here, we abbreviate $f'  := \frac{\dd f}{\dd x}$, etc.
We now use two different types of functions for $f$ and $g$: on the
one hand, we keep the eigenvalues by conjugation with unitary matrices, 
on the other hand, we modify them by multiplication with 
appropriate commuting matrices. 
Namely, let first 
\bgl
f(x) & := & \e^{- x C} A \e^{x C} \:\:\: \text{ for $C \in \su(n)$,} 
\egl\noindent
i.e., $C^\ast = -C$ and $\trm C = 0$.
Then $f(0) = A$ and $f'(0) = [A,C]$. Moreover, $f(x)$ is $n$-phone for 
any $x$ and any $n$-phone $A$.
If now $(A,B)$ is an extremal point for $\trm S_{m,k}$
among the positive matrices with unit $\nl$-norm,
then we have for all $C \in \su(n)$
\bgl
\hspace*{-\parindent}
0 & = & \trm S'_{m,k} (f,B) \einschr{x=0} 
  \breitrel= m \: \trm \bigl([A,C] \: S_{m-1,k}(A,B)\bigr) .
\egl\noindent
Since $A$ and $S_{m-1,k}(A,B)$ are hermitian \cite{bmv16},
we get $[S_{m-1,k}(A,B),A] = \NULL$ from Lemma \ref{lem:killingform}.
Now, secondly, we consider
\bgl
f(x) & := & \frac{A \e^{x C}}{\norm{A \e^{x C}}_\nl} \:\:\: \text{ for $C \in \gl(n)$.} 
\egl\noindent
Note that $f$ may fail to be differentiable at $x = 0$ for $\nl = \infty$.
In fact, let $E_{ij}$ be the matrix having entry $1$ at position $(i,j)$ 
and zeros elsewhere. Consider $A := E_{11} + E_{22}$ and
$C := E_{11}$. Then $\norm{A \e^{x C}}$ equals $\e^x$ for $x \geq 0$ and
$1$ for $x \leq 0$, which is obviously not differentiable.
In general, the problem arises if the maximal eigenvalue of $A$ is
of multiplicity $2$ or higher. Therefore, for the moment, we assume $\nl$
to be finite.
One easily%
\footnote{
Observe that
\bgl[2ex]
\norm{A \e^{xC}}'_\nl (0) 
 & = & \inv{\nl} \norm{A}_\nl^{1-\nl} \bigl(\norm{A \e^{xC}}_\nl^\nl\bigr)' (0)
 \breitrel= \inv{\nl} \norm{A}_\nl^{1-\nl} \bigl(\trm (A \e^{xC})^\nl \bigr)' (0) 
 \breitrel=
            \norm{A}_\nl^{1-\nl} \: \trm A^\nl C
\egl\noindent
and, therefore,
\bgl[2ex]
f'(0) 
  & = & \inv{\norm{A}_\nl^2} 
        \bigl(AC \norm{A}_\nl - A \norm{A}_\nl^{1-\nl} \: \trm A^\nl C \bigr)
  \breitrel= \inv{\norm{A}_\nl^{\nl+1}} 
        \bigl(AC \norm{A}^\nl_\nl - A \: \trm A^\nl C \bigr).
\egl

}
 checks that
\bgl[2ex]
f'(0) 
  \breitrel= \inv{\norm{A}_\nl^{\nl+1}} \bigl(AC \: \trm A^\nl - A \: \trm A^\nl C \bigr).
\egl\noindent
Of course, a priori, it is not clear that $f(x)$ is positive and hermitian,
even for small $x$. But, if $U$ is some unitary matrix,
such that $U^\ast A U$ (and $U^\ast S_{m-1,k}(A,B)U$) is diagonal, 
then $f$ is positive and hermitian
for any $C = U D U^\ast$ with $D$ being diagonal and real. In fact, the product of
diagonal positive and hermitian matrices has these properties again.
If now $A$ and $B$ are nonzero and again extremal for $\trm S_{m,k}$
among $n$-phone matrices, then
\bgl
\hspace*{-\parindent}
0 \breitrel= \frac{\norm{A}_\nl^{\nl+1}}m \: \trm S'_{m,k} (f,B) \einschr{x=0} 
  & = & \trm \bigl((AC \: \trm A^\nl - A \: \trm A^\nl C ) S_{m-1,k} (A,B) \bigr)\\
  & = & \trm \bigl(S_{m-1,k} (A,B) A \: \trm A^\nl - A^\nl \: \trm S_{m-1,k} (A,B) A \bigr) C \:.
\egl\noindent
Since $S_{m-1,k}(A,B)$ and $A$ commute as seen above and are 
hermitian, we get
\bgl
 S_{m-1,k} (A,B) A \: \trm A^\nl & = & A^\nl \: \trm S_{m-1,k} (A,B) A
\egl\noindent
from Lemma \ref{lem:spurform}.
Similarly, we can derive 
\bgl
 S_{m-1,k-1} (A,B) B \: \trm B^\nl & = & B^\nl \: \trm S_{m-1,k-1} (A,B) B.
\egl\noindent
Altogether we have
\bprop
If $0 < k < m$ and if $\trm S_{m,k}$ is extremal at $(A,B)$
for the positive hermitian matrices 
having unit $\nl$-norm with $1 \leq \nl < \infty$,
then
\bgl
 S_{m-1,k} (A,B) A & = & A^\nl \: \trm S_{m-1,k} (A,B) A \\
 S_{m-1,k-1} (A,B) B & = & B^\nl \: \trm S_{m-1,k-1} (A,B) B \:.
\egl\noindent
\eprop
The case $\nl = 2$ has already been derived by Hillar in \cite{bmv16}.
There, the norm equals the Fro\-be\-nius norm.
The case $\nl = \infty$, i.e., the supnorm case, can be dealt with
as for $\nl < \infty$ as far as we derive that $A$ and $S_{m-1,k}(A,B)$ commute.
Assuming now, for simplicity, that $A$ and $S_{m-1,k}(A,B)$ are diagonal
and $\norm A = 1$, we see that $f(x) := A \e^{xC}$ is (at least for
small $\betrag x$) $n$-phone --- provided $C$ is diagonal with $C_{ii} = 0$ for $A_{ii} = 1$.
Then $f'(0) = AC$ implying $\trm ACS_{m-1,k}(A,B) = 0$,
whence the $(i,i)$ components of $S_{m-1,k}(A,B) A$ vanish if 
$A_{ii} \neq 1$. If $P_A$ has a single nonzero entry, we immediately
get $S_{m-1,k} (A,B) A = P_A \: \trm S_{m-1,k} (A,B) A$.
In the other case, however, we run into the non-differentiability problem
as above. At present, we are not able to solve this problem.

Nevertheless, we have
\bcorr
If $0 < k < m$ and if $\trm S_{m,k}$ is extremal at $(A,B)$
for the positive hermitian matrices 
having unit $\nl$-norm with $1 \leq \nl < \infty$,
then
\bgl
 S_{m,k} (A,B)
  & = & \frac{(m-k) A^\nl + k B^\nl}m \: \: \trm S_{m,k} (A,B) \:.
\egl\noindent
The same is true for $\nl = \infty$, provided $P_A$ and $P_B$ 
have rank $1$.
\ecorr
\bpf
Use 
the properties of Hurwitz products listed in Subsection \ref{subsect:hurwitzproducts}.
\qed
\epf

\neueseite

\section{Lie Algebra Relations}

\blem
\label{lem:killingform}
If $A$ and $S$ are hermitian matrices, fulfilling $\trm [A,C] S = 0$
for all $C \in \su(n)$, then $A$ and $S$ commute.
\elem
\bpf
Since $A$ and $S$ are hermitian, we have 
$[A,S]^\ast = -[A,S]$
and, anyway, $\trm[A,S] = 0$. Therefore, $[A,S] \in \su(n)$.
Moreover, $\trm [S,A] C = \trm[A,C] S$ vanishes by assumption 
for any $C \in \su(n)$.
Since $\su(n)$ is semisimple, 
the Killing form $(X,Y) := \inv n \: \trm (X Y)$ 
on $\su(n)$ is 
non-degenerate, giving $[S,A] = \NULL$.
\qed
\epf

\blem
\label{lem:spurform}
If $A$ and $S$ are hermitian matrices
that are diagonal after conjugation with $U$
and fulfill
$\trm \bigl((S A \: \trm A^L - A^L \: \trm S A) U D U^\ast \bigr) = 0$
for all diagonal matrices $D$, then $S A \: \trm A^L = A^L \: \trm S A$.
\elem
\bpf
If $A$ and $S$ are already diagonal, then the assertion is 
trivial. In fact, letting $D$ be the matrix having just a single nonzero
entry at position $(i,i)$, the trace equation above means that 
the $(i,i)$ component of $(S A \: \trm A^L - A^L \: \trm S A)$ vanishes.
Since the off-diagonal elements are zero anyway, we get the assertion.

In the general case observe that 
\bgl
 \trm \bigl(U^\ast S U \: U^\ast A U \: \trm (U^\ast A U)^L 
          - (U^\ast A U)^L \: \trm U^\ast S U \: U^\ast A U\bigr) D \hspace*{-18ex}
\\
  & = & \trm \bigl(S A \: \trm A^L - A^L \: \trm S A \bigr) U D U^\ast 
\egl   
reduces this case to the first one.
\qed
\epf


\section{Simple, But Useful Identities}
\label{app:useful_identity}
\blem
\label{lem:prod_diff_zerlegung}
For any $n \kreuz n$ matrices $X_i$ and $X$, we have
\bgl
X_1 \cdots X_k
 & = & X^k + \sum_{i=1}^{k} 
        X^{i-1} \: (X_i - X) \: X_{i+1} \cdots X_k.
\egl
\elem
\bpf
For $k = 1$, we have 
$X_1 = X^1 + X^{0} \: (X_1 - X)$.
For $k > 1$, we have by induction
\bgl
X_1 \cdots X_{k+1}
 & = & X^k X_{k+1} + \sum_{i=1}^{k} 
        X^{i-1} \: (X_i - X) \: X_{i+1} \cdots X_k X_{k+1} \\
 & = &  X^k \bigl(X + (X_{k+1} - X)\bigr) + \sum_{i=1}^{k} 
        X^{i-1} \: (X_i - X) \: X_{i+1} \cdots X_k X_{k+1} \\
 & = &  X^k X + \sum_{i=1}^{k+1} 
        X^{i-1} \: (X_i - X) \: X_{i+1} \cdots X_k X_{k+1}.
\egl
\qed
\epf

\blem
\label{lem:eulerseries}
For any natural number $k$ and for all $\betrag\tau < 1$, we have 
\bgl
\inv{(1-\tau)^{k+1}} & = & \sum_{\laufm = 0}^\infty \binom{k+\laufm}{\laufm} \: \tau^\laufm\,.
\egl
\elem
\bpf
The statement is clear for $k = 0$. Now, we have inductively
\bgl
\frac{k}{(1-\tau)^{k+1}} 
  & = & \Bigl(\inv{(1-\tau)^k}\Bigr)' 
  \breitrel= \sum_{\laufm = 1}^\infty \binom{k-1+\laufm}{\laufm} \: \laufm \tau^{\laufm-1} \\
  & = & \sum_{\laufm = 0}^\infty \binom{k+\laufm}{\laufm+1} \: (\laufm + 1) \tau^{\laufm}
  \breitrel= \sum_{\laufm = 0}^\infty \binom{k+\laufm}{\laufm} \: k \tau^{\laufm} \,.
\egl
\vspace*{-\baselineskip}\qed
\epf

\end{document}